\documentclass[graybox]{svmult}

\usepackage{mathptmx}       % selects Times Roman as basic font
\usepackage{helvet}         % selects Helvetica as sans-serif font
\usepackage{courier}        % selects Courier as typewriter font
\usepackage{type1cm}        % activate if the above 3 fonts are
\usepackage{makeidx}         % allows index generation
\usepackage{graphicx}        % standard LaTeX graphics tool
\usepackage{multicol}        % used for the two-column index
\usepackage[bottom]{footmisc}% places footnotes at page bottom
\usepackage{hyperref}        %for hyperlinks
\usepackage{soul}            % for high-lighting of text
\usepackage{color}

\usepackage[dvipsnames]{xcolor}
\definecolor{darkblue}{rgb}{0,0.1,0.55} 
\hypersetup{
    colorlinks = true,
    allcolors = darkblue,
}

\usepackage[square,numbers]{natbib}
\makeindex 

\usepackage{amsmath}
\usepackage{amssymb}

\renewcommand{\d}{\mathrm{d}}
\newcommand{\pder}[3][]{\frac{\partial^{#1}{#2}}{\partial{#3}^{#1}}}
\newcommand{\ix}[1]{#1}

\begin{document}
%\tableofcontents{}
\title*{Non-linear effects in EMRI dynamics and their imprints on gravitational waves}
\author{Georgios Lukes-Gerakopoulos\thanks{corresponding author},  Vojt\v{e}ch Witzany}
\institute{Georgios Lukes-Gerakopoulos \at Astronomical Institute of the Czech Academy of Sciences, Bo\v{c}n\'{i} II 1401/1a, CZ-141 00 Prague, Czech Republic. \email{gglukes@gmail.com}
\and Vojt\v{e}ch Witzany \at School of Mathematics and Statistics, University College Dublin, Belfield, Dublin 4, D04 V1W8, Ireland. \email{vojtech.witzany@ucd.ie}
}
\maketitle
\abstract{ The largest part of any gravitational-wave inspiral of a compact binary can be understood as a slow, adiabatic drift between the trajectories of a certain referential conservative system. In many contexts, the phase space of this conservative system is smooth and there are no ``topological transitions'' in the phase space, meaning that there are no sudden qualitative changes in the character of the orbital motion during the inspiral. However, in this chapter we discuss the cases where this assumption fails and non-linear and/or non-smooth transitions come into play. In integrable conservative systems under perturbation, topological transitions suddenly appear at resonances, and we sketch how to implement the passage through such regions in an inspiral model. Even though many of the developments of this chapter apply to general inspirals, we focus on a particular scenario known as the Extreme mass ratio inspiral (EMRI). An EMRI consists of a compact stellar-mass object inspiralling into a supermassive black hole. At leading order, the referential conservative system is simply geodesic motion in the field of the supermassive black hole and the rate of the drift is given by radiation reaction. In Einstein gravity the supermassive black hole field is the Kerr space-time in which the geodesic motion is integrable. However, the equations of motion can be perturbed in various ways so that prolonged resonances and chaos appear in phase space as well as the inspiral, which we demonstrate in simple physically motivated examples.
}

%Each chapter should be preceded by an abstract (about 250 words) that summarizes the content. The abstract will appear \textit{online} at \url{www.SpringerLink.com} and be available with unrestricted access. This allows unregistered users to read the abstract as a teaser for the complete chapter. Please do not include reference citations, cross-references or undefined abbreviations in the abstract.}
\section*{Keywords} 
black holes, LISA, EMRI, celestial mechanics, chaos, dynamical systems 

\section{Introduction}

An \ix{Extreme Mass Ratio Inspiral} (EMRI) is an event that is expected to occur once a stellar-mass compact object (the secondary) is captured on a sufficiently tight orbit by a supermassive black hole (the primary) in a center of a galaxy. As in the case of any compact-object binary, the motion of the two bodies in this specific binary creates gravitational waves that carry energy and angular momentum to infinity. Consequently, the orbit decays in a spiralling motion and the stellar-mass compact object eventually plunges into the supermassive black hole. The gravitational waves carrying the imprint of this motion peak in the mHz frequency band and are expected to be observed by future gravitational-wave observatories such as the Laser Interferometer Space Antenna (LISA) \cite{LISA}. However, in contrast to stellar-mass compact binaries observed by the ground-based detectors LIGO and Virgo, a sizable fraction of EMRIs should enter the sensitivity band in and eccentric, orbitally precessing state. As a result, we have to consider EMRI scenarios from a richer and more complex landscape of dynamics.

A common approximative framework for the description of this special class of gravitational-wave inspirals is that the secondary is replaced by a point particle\footnote{The correspondence between the position of the ``particle'' and the real body is established through matched asymptotic expansions \cite{poisson2011}, see also Chapter \ref{chap:self-force} \textit{Black hole perturbation theory and gravitational self-force} in this handbook. Roughly speaking, the position of the particle corresponds to the center of mass of the body.} in the field of the much more massive primary with equations of motion ordered by the mass ratio $q=\mu/M$, where $\mu$ is the mass of the secondary and $M$ the mass of the primary. At zeroth order the equations of motion are those of a free test particle, or a geodesic, in the field of the primary black hole. First and second order corrections are then ``self-force terms'' obtained from black-hole perturbation theory as well as effects due to the finite size of the body \cite{Barack19,Witzany19b}. The corrections to the zeroth-order equations then cause small local deviations from geodesic motion but the most important global effect is the gradual decay of the orbit.

Geodesic motion in the field of an isolated spinning black hole in Einstein gravity, the Kerr space-time, is integrable \cite{Carter68}. Consequently, if we only model an EMRI as motion adiabatically drifting from geodesic to geodesic in Kerr space-time, we obtain a reasonably simple waveform with slowly drifting fundamental frequencies and harmonics. Will this simplicity hold once we refine the inspiral model? Unfortunately no, at least generically. Once we include immediate, ``conservative'' effects to the equations of motion or introduce even slight modifications to Kerr space-time, the integrability is mostly broken. Discussing the precise properties of the resulting near-integrable of weakly chaotic systems and the consequences for the inspiral are the subjects of this chapter. 

\begin{figure}
    \centering
    \includegraphics[width=\textwidth]{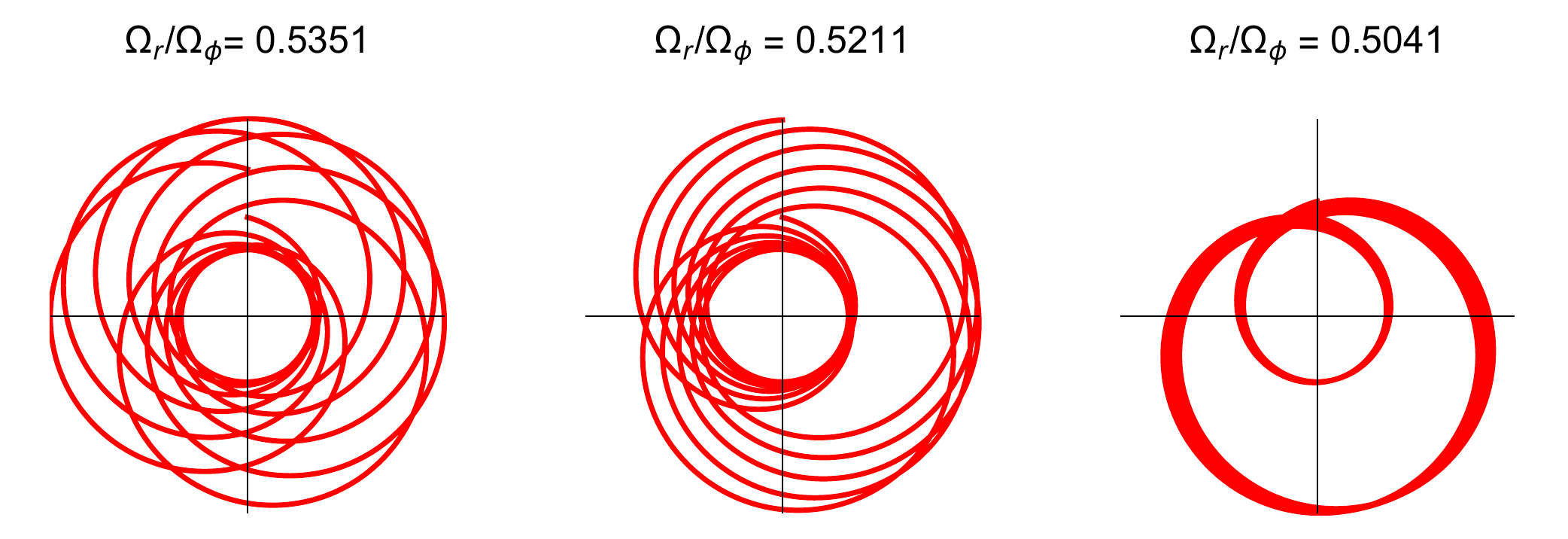}
    \caption{Three orbits with varying ratios of the radial oscillation frequency $\Omega_r$ and azimuthal rotation frequency $\Omega_\phi$ plotted for ten azimuthal cycles. The leftmost, almost generic orbit very quickly samples all the possible phase shifts between the $r$ and $\phi$ motion and densely covers the available space. However, as the ratio $\Omega_r/\Omega_\phi$ approaches close to $1/2$, the orbit is averaging over the available space at a slower and slower rate, which can even be slower than the inspiral itself.}
    \label{fig:rezoorb}
\end{figure}

In an EMRI system, chaos itself is not expected to be a prominent effect, even if an EMRI can pass through a chaotic layer for an extremely brief period of time. On the other hand, passage through resonant regions in an EMRI pose the biggest challenge for waveform models. At this point we should define what do we even mean by the word \ix{\textit{resonance}}. The broadest definition of a resonance is when two or more characteristic frequencies of a system match in integer ratios. This means that there is a relative phase of the motions that is ``frozen by kinematic coincidence'' and the resonant orbit stops sampling the available phase space (see Fig. \ref{fig:rezoorb}). This has more than one consequence, which has lead to some confusion in the literature. 

An obvious issue is that an orbit sampling only a part of the dynamically available phase space may not have the symmetries of the equations of motion. Hence, even though the equations of motion may posses rotational symmetry, the resonant orbit may not. In the context of gravitational radiation this can mean that it radiates anisotropically, which results in a resonantly enhanced kick to the binary in question \citep[see][]{maarten2014}. In many gravitational-wave applications one is interested only in the time-averaged flux of gravitational waves over the orbit and for non-resonant orbits the time average is interchangeable with the more convenient phase-space average. However, the resonant orbits evolve through a smaller subset of the phase space, which means the averaging formulas have to be modified for them in such computations \citep{isoyama2013,isoyama2019}.

Nevertheless, the subject of this chapter is different. We are concerned with the fact that an integrable, conservative dynamical system under perturbation often develops a special dynamical phenomenon \textit{in a region of non-zero phase-space volume} around the original kinematic resonances, which we will call a \textit{prolonged resonance} \cite{LGAC10,Zelenka20}. A prolonged, or perhaps ``inflated'' resonance is a region in phase space where trajectories are observed to \textit{oscillate} around a finite subset of the original resonant trajectories. The qualitative transition between generic motion and the prolonged resonance has to be treated with special care in an inspiral computation. 

In the EMRI literature, one can encounter the terms \textit{transient} and \textit{sustained} resonances during an inspiral \cite{Flanagan12,vandeMeent14}. What is the relation of these terms to our ``\ix{prolonged resonance}''? The self-force on the inspiraling body in the EMRI can be formally decomposed into parts that cause the secular decay of the orbit and those which do not (conventionally called ``averaged dissipative self-force'' and ``oscillating dissipative and conservative self-force'' respectively). Taking only the second part into account, we obtain a virtual conservative dynamical system that has the character of some sort of perturbed geodesic motion in Kerr space-time. Even though it has been posited that even this perturbed system could stay integrable \citep{Flanagan12}, we generically expect that it will not be and that it will contain prolonged resonances. Depending on the character of both the prolonged resonances and the secular part of the self-force, the encounter of the real inspiral with these structures can be transient or, under special conditions, sustained for a time comparable with the inspiral time. That is, transient and sustained resonances as referred to in the EMRI literature can be understood as different modes of interaction with the topological structure of the prolonged resonance.

In this chapter, we first establish the general mathematical theory of dynamical systems, focusing on perturbed Hamiltonian systems. The center stage is occupied by prolonged resonances and we also sketch how to treat the inspiral through it. After establishing the general theory and tools, we pass to specific cases of near-integrable systems and provide some numerical examples. 

Throughout the chapter we use the $G=c=1$ geometric units, the $(-,+,+,+)$ signature of the metric, and bold characters denote vectors ($n$-tuples). From Sec.~\ref{sec:OrbKerr} on the Einstein summation convention is employed and the Greek indices $\mu,\nu,\kappa,\lambda,...$ run from $0$ to $3$.

\section{Brief introduction to dynamical systems}

\subsection{Continuous and discrete dynamical systems}
A system evolved forward in time by a set of equations is a \ix{dynamical system}. Depending on whether the time evolution is taking place in continuous or discrete time steps, the dynamical systems split into continuous systems and mappings respectively. Formally, a set of first order differential equations 
\begin{align}\label{eq:cds}
    \frac{\d \mathbf{x}}{\d t}=\mathbf{f}(\mathbf{x},t)
\end{align}
defines a \ix{{\em continuous dynamical system}}. In Eq.~\eqref{eq:cds} $\mathbf{x}$ is a vector belonging to the phase space $\mathcal{S}$, $\mathbf{f}$ is a vector function on the phase space, and $t$ is a continuous evolution variable, which usually has the meaning of time. For the purposes of our discussion, we always assume that $f$ is sufficiently smooth. We say that the phase-space has dimension $D$ if the phase-space vector $\mathbf{x}$ has dimension $D$. Equation~\eqref{eq:cds} is called the equation of motion and its solution with respect to specific initial conditions is called a trajectory in phase space. On the other hand, a set of difference equations 
\begin{align}\label{eq:maps}
   \mathbf{x}_{n+1}=\mathbf{F}(\mathbf{x}_n,n)
\end{align}
defines a {\ix{\em discrete mapping}}. In  Eq.~\eqref{eq:maps} $\mathbf{x}_{n}$ is again a vector in phase space, $\mathbf{F}$ is a vector function and $n\in\mathbb{N}$ is a label for the discrete time steps. 

\begin{figure}
    \centering
    \includegraphics[width=0.7\textwidth]{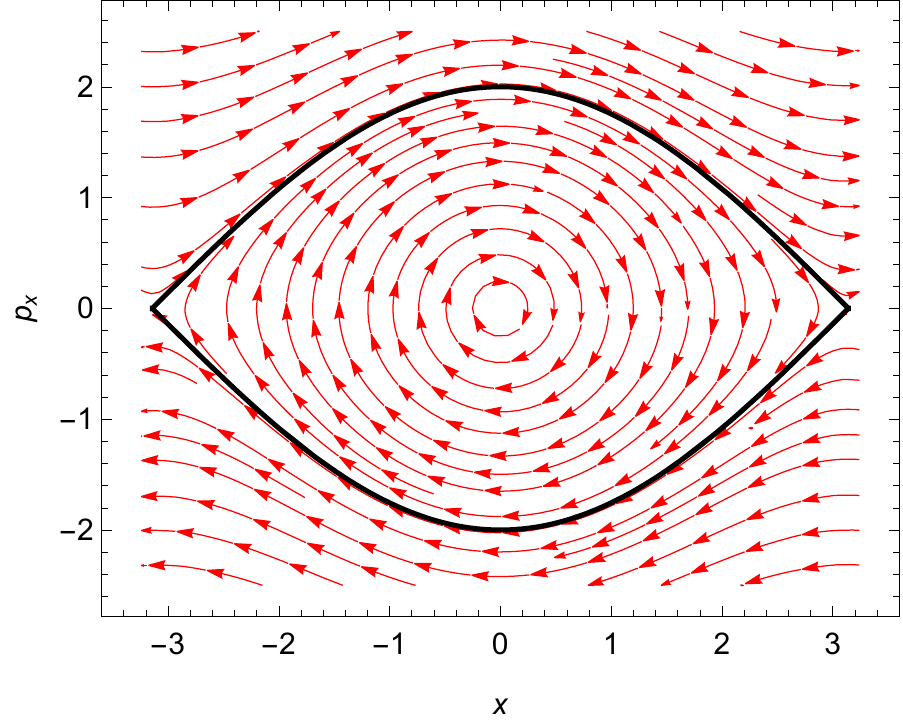}
    \caption{The phase-space portrait of a non-linear pendulum with the Hamiltonian $H = p_x^2/2 - \cos(x)$. The arrows denote the direction of the flow and the black line is the separatrix $H = 0$ separating the topologically distinct oscillations around $x=0$ (also known as librations) from the rotations (in the sense that the motion continuously rotates through the periodic $x\in (0, 2\pi]$). Both the types of motion can be understood as bound and the phase-space trajectory a topological circle ($\mathbb{T}^1 = S^1$), but they belong to a different homotopy class on the phase-space manifold.}
    \label{fig:NLpendulum}
\end{figure}

In the framework of General Relativity the dynamical systems in question are continuous. Even though continuous systems can be reduced to mappings, as we will discuss later on, for now we are going to focus only on the former. In continuous dynamical systems the equations of motion define a flow in phase space $\mathcal{F}_t:\mathcal{S}\rightarrow \mathcal{S}$ along which an initial condition $\mathbf{x}_0$ evolves to $\mathbf{x}$ in time $t$, i.e. $\mathbf{x}(t)=\mathcal{F}_t(\mathbf{x}_0)$. If there is a volume element on $\mathcal{S}$ such that the size of any volume of initial conditions along the flow does not change, then the system is called {\ix{\em conservative}}. If  $\mathbf{f}$ in Eq.~\eqref{eq:cds} does not depend explicitly on time, then the system is {\ix{\em autonomous}}. An example of a phase-space flow of an autonomous conservative system with a phase space of dimension 2 is given in Fig. \ref{fig:NLpendulum}.

\subsection{Hamiltonian systems and integrability}

Consider a Hamiltonian system with $N$ degrees of freedom, that is, a phase space consisting of $N$-dimensional positional vector $\vec{q} = (q^1,...,q^n)$ and a corresponding set of $N$ conjugate momenta $\vec{p} = (p_1,...,p_N)$. Hamilton's equations then read
\begin{align} 
\frac{\d \mathbf{q}}{\d t}=\frac{\partial H}{\partial \mathbf{p}},\quad \frac{\d \mathbf{p}}{\d t}=-\frac{\partial H}{\partial \mathbf{q}} \, .
\end{align}
These can be put in the form
\begin{align}\label{eq:HamCont}
  \frac{\d \mathbf{x}}{\d t}=
    \begin{bmatrix}
     0 & I_N  \\
    -I_N & 0 
\end{bmatrix}
\cdot \frac{\partial H}{\partial \mathbf{x}}\,=\mathbf{f}(\mathbf{x}) ,
\end{align}
where $I_N$ is the $N$-dimensional identity matrix and $\vec{x} = (\vec{q},\vec{p})$. In other words, a Hamiltonian system with $N$ degrees of freedom is a dynamical system with a phase space of dimension $2N$. Hamilton's equations can be shown to conserve the volume form\footnote{An even stronger statement can be proven, the equations can be shown to conserve the so-called symplectic differential form $\mathrm{d}\vec{p} \wedge \mathrm{d} \vec{q}$ from which the conservation of the volume form trivially follows.} $\mathrm{d}^N\vec{p} \,\mathrm{d}^N\vec{x}$, which means Hamiltonian systems are conservative \cite{Poincare93}.

The Hamiltonian system is autonomous when ${\partial H}/{\partial t}=0$, which also implies $\d H/\d t = 0$ along the trajectories. In other words, in autonomous systems $H$ is a constant of motion and its value depends only on the initial condition $x_0$. We say that any differentiable function $I(\mathbf{x})$ is a \ix{{\em constant of motion}} (or an {\em integral of motion}) iff ${\d \, I}/{\d t}=0$ for any trajectory in $\mathcal{S}$. 

An autonomous Hamiltonian system is called {\ix{\em integrable}} iff it has at least as many constants of motion as degrees of freedom and this set of constants of motion is functionally independent and in involution. Two functions on phase-space $I_1(\vec{x}),~I_2(\vec{x})$ are said to be in involution if 
\begin{align}
    \{I_1,I_2\}= \frac{\partial I_1}{\partial \vec{x}} \cdot \begin{bmatrix}
     0 & I_N  \\
    -I_N & 0 
\end{bmatrix} \cdot \frac{\partial I_2}{\partial \vec{x}} = 0\,,
\end{align} 
where  $\{~\}$ is the Poisson bracket. In that case the trajectories stay on the hypersurface defined by $I_i = {\rm const.},\, i=1,...,N$. Such a set of conditions defines an $N$-dimensional manifold $\mathcal{M}$, which is smooth and invariant under the action of the flow $\mathcal{F}_t$ in $\mathcal{S}$. If the manifold $\mathcal{M}$ is compact and connected, then according to the Liouville-Arnold theorem $\mathcal{M}$ is diffeomorphic to the $N$-dimensional torus $\mathbb{T}^N$. This case corresponds to {\em bound integrable motion}. Consequently, one can transform to a convenient set of canonical variables known as \ix{{\em action-angle coordinates}} $(\pmb{\theta},\mathbf{J})$ such that $H(\mathbf{x},\mathbf{p}) = H(\mathbf{J})$ and the equations of motion reduce to
\begin{align}
    \dot{\pmb{\theta}} \equiv \pmb{\omega}(\mathbf{J}) = \frac{\partial H}{\partial \mathbf{J}}\,, \;\dot{\mathbf{J}} = - \frac{\partial H}{\partial \pmb{\theta}} = 0\,.
\end{align}
The $\pmb{\theta}$ variables correspond to the angles on the torus  $\mathbb{T}^N$, while their conjugate momenta, ``the actions'', $\mathbf{J}$ correspond to the integrals of motion.

If we specify all of the $\mathbf{J}$, then both the torus on which the motion takes place as well as the fundamental frequencies of motion $\pmb{\omega}(\mathbf{J})$ are also specified. We can vary $\mathbf{J}$, to explore the \ix{\textit{foliation of the phase space}} $\mathcal{S}$ by the tori. On the other hand, if the set of frequencies $\pmb{\omega}$ is to uniquely specify the torus on which we are moving, the following non-degeneracy condition must also hold globally
\begin{align}\label{eq:nondeg}
 \det\left(\frac{\partial \pmb{\omega}}{\partial \mathbf{J}}\right)\neq0 \,.    
\end{align}
In particular, this condition is violated for harmonic motion and near non-degenerate equilibrium points where $\pmb{\omega}$ is locally a constant vector. In other words, we cannot  tell the values of actions (which are proportional to oscillation amplitudes) from the frequencies of near-equilibrium oscillations alone. 

The character of the motion on the torus depends on whether any of the fundamental frequencies match in an integer ratio. More generally, it depends on whether there exist linearly independent integer-vectors $\mathbf{k}$ such that
\begin{align} \label{eq:res_cond}
    \mathbf{k}\cdot \pmb{\omega} \equiv \sum_{i=1}^N k_i \omega^i=0,\: \textrm{where} \: k_i\, \in\, \mathbb{Z} \:\: \textrm{and} \:\: |\mathbf{k}|\equiv\sum_{i=1}^N |k_i|\neq 0 \, .
\end{align}
Eq.~\eqref{eq:res_cond} is called a \ix{{\it resonance condition}}. The number of linearly independent $\mathbf{k}$ for which this holds true is called the number of resonant conditions fulfilled by the motion. The motion on a torus $\mathbb{T}^N$ is \ix{{\em quasiperiodic}} if no resonant condition is fulfilled. A quasiperiodic orbit will densely cover the torus in infinite time and it will not return to the initial condition from where it started in finite time. On the other hand, the motion is also \ix{\textit{ergodic}}, which means that an infinite-time average of a phase-space function along the motion can be replaced by a phase-space average over the torus.

If there are $m<N-1$ independent resonant conditions, then the quasiperiodic orbit will cover densely a $\mathbb{T}^{N-m}$ torus which is a submanifold of the respective $\mathbb{T}^N$ torus. If there are  $m=N-1$ resonant conditions, then the motion is \ix{{\it periodic}}. By solving the respective system of resonant conditions in the fully periodic case we can pick one of the frequencies, e.g. $\omega^1$, and express all the other frequencies as $\omega^i=r_i \omega^1$, where $r_i\in \mathbb{Q}$. 

\subsection{Poincar{\'e} surfaces of section}

\begin{figure}
    \centering
    \includegraphics[width=\textwidth]{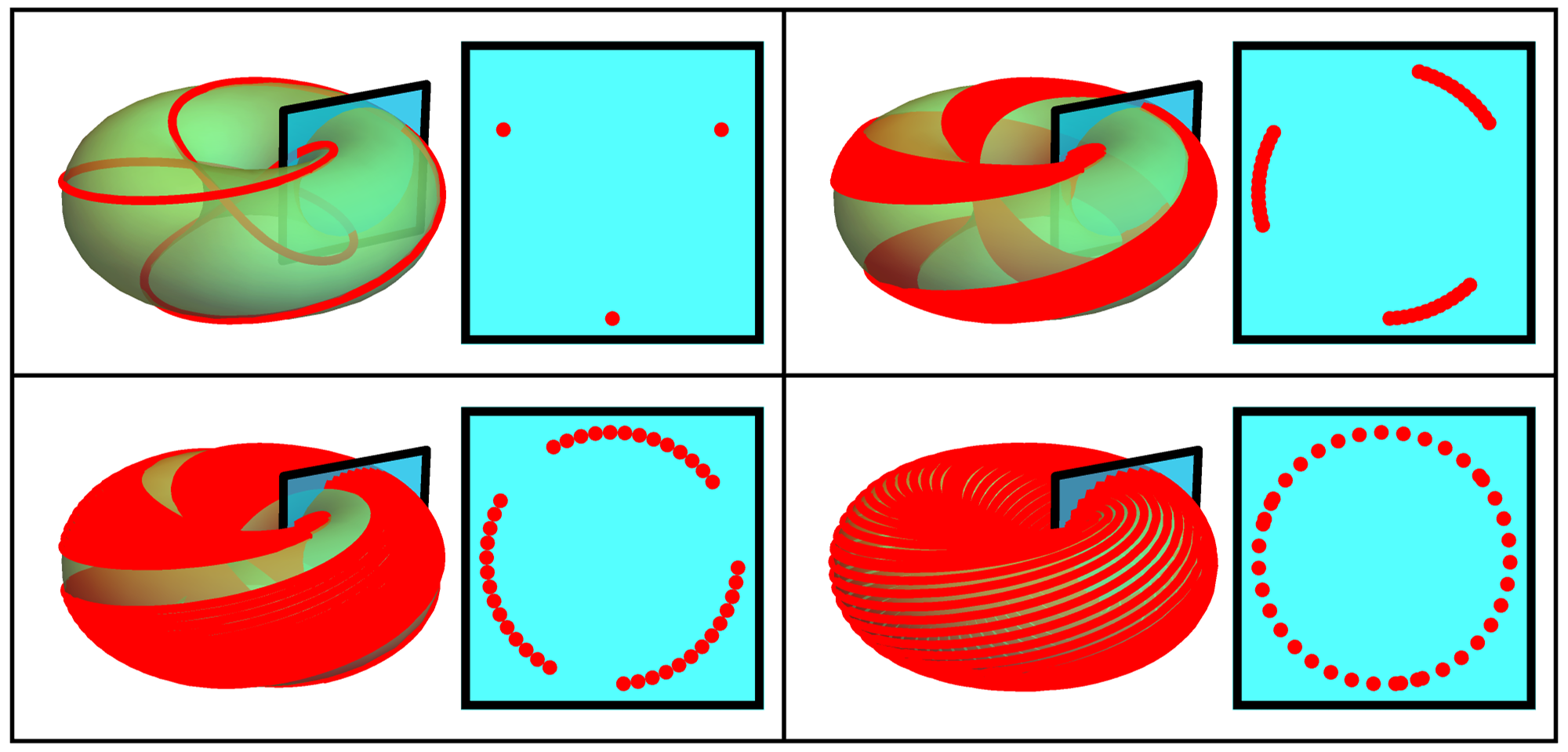}
    \caption{An illustration of sections obtained from 40 intersections of various phase-space trajectories (red) through a Poincaré surface of section (cyan). The trajectories stay on the torus defined by the integrals of motion (green). The top left trajectory is a 3:2 resonant trajectory (the frequency of motion over the small circle of the torus is two thirds of that over the large circle) and it repeatedly intersects only a set of three points on the section. The rest of the trajectories are non-resonant and they asymptotically trace out a cross-section of the torus.}
    \label{fig:sectionIll}
\end{figure}

Let us now discuss the visualisation of the foliation of tori in autonomous Hamiltonian systems of 2 degrees of freedom. Such dynamical systems have 4-dimensional phase spaces and $\mathbb{T}^2$ tori foliating them, depending on two integrals of motion (one of the integrals can be chosen to be the Hamiltonian $H$). If we restrict ourselves to a hypersurface where one of the integrals of motion is kept fixed, we are reduced to a 3-dimensional space filled with 2-dimensional tori, which is still hard to visualize. Thus, we have to make a well-chosen section through this space that transversely (non-tangentially) cuts through the tori and allows us to examine the foliation (see Fig. \ref{fig:sectionIll}). This is a so-called \ix{{\em Poincar\'{e} surface of section}}, which actually corresponds to a discrete mapping, called \textit{Poincar{\'e} mapping}, as defined in equation \eqref{eq:maps}, because every consecutive point is uniquely determined by the previous one.

To construct the surface of section\footnote{The term Poincar\'{e} surface of section is most often reduced to just Poincar\'e section or surface of section in the bibliography. Following this tradition, we use these reduced terms in the article interchangeably.} numerically, one has to integrate the equations of motion and identify the constant of motion to be held fixed (often the Hamiltonian) and a good section condition $\Phi(\vec{x},\vec{p}) = 0$. For instance, if we know that all orbits of interest oscillate about a certain equilibrium point, it is good to put the section into that point. Once the trajectory passes through $\Phi(\vec{x},\vec{p}) = 0$, the only two remaining phase-space coordinates are recorded and plotted, and this is repeated until many points from a single trajectory are gathered. Since in the integrable case a quasi-periodic trajectory densely fills the torus, the set of points from it gradually circle out a single closed curve on the plot ant it is called an invariant curve, since on a Poincar\'{e} section it maps itself on itself. As this is repeated with a number of independent trajectories, the foliation is revealed as a set of nested non-intersecting closed curves (see Fig. \ref{fig:sec_rot}). On the other hand, trajectories fulfilling a resonant condition only fill a subspace of the torus and appear as a finite set of periodically repeating points on the section. For each resonance the number of these sets is infinite. Each set, however, consists of a finite number of points equal to the {\em periodicity} of the resonance\footnote{The periodicity of the resonance is also called the multiplicity of the resonance.}, i.e. to the number of mappings needed for a periodic orbit to return to its initial condition on the Poincar\'{e} section. The applicability of Poincar{\'e} surfaces of section goes well beyond bound integrable systems. On a surface of section one can observe the breaking of integrability, mainly by studying the neighborhood of periodic points as explained in section~\ref{sec:stab}. 

The case of two degrees of freedom is the lowest number of degrees of freedom in which non-integrability can occur, and it is thus also the best studied case. On the other hand, most physical systems of interest have more degrees of freedom, so it would seem the Poincar{\'e} surface of section is not useful for them. In that case, one has to carefully consider the symmetries of the problem and see whether the essential dynamics can be observed only in a sub-sector of the system, i.e. in a \ix{{\it reduced system}}. For example, motion of a particle in a 3-dimensional axisymmetric potential has 3 degrees of freedom, but the azimuthal symmetry angle is redundant and the azimuthal angular momentum $L_z$ is constant along orbits. We can then understand the motion of all particles with the same $L_z$ as a family of dynamical systems of two degrees of freedom where $L_z$ plays the role of a parameter of the system.

\subsection{Stability of orbits in maps and continuous systems} \label{sec:stab}

Periodic and fixed points can be considered as the skeleton of a dynamical system, since they are tied to ``topological transitions'' in the flow or, in physical terms, to qualitative changes of the motion in phase space. Their stability can be found by applying linear perturbation theory. This can be done in the sense of a discrete dynamical system on a Poincar{\'e} surface of section or for the continuous flow in the full phase space, both of which will now be discussed.

\subsubsection{Fixed points in discrete dynamical systems} \label{sec:discstab}
Let $\mathbf{x}_{\rm f}$ be a fixed point of the mapping $\mathbf{F}(\mathbf{x}_n)$, i.e. $\mathbf{x}_{\rm f}=\mathbf{F}(\mathbf{x}_{\rm f})$. For every periodic point $\mathbf{x}_{\rm p}$ with periodicity $j$ a new mapping $\mathbf{H}=\mathbf{F}^j(\mathbf{x}_n)$ can be defined so that $\mathbf{x}_{\rm p}=\mathbf{H}(\mathbf{x}_{\rm p})$. Thus, we can implicitly treat both periodic and fixed points by only discussing fixed points. The stability of a fixed point can be examined by a linear perturbation $\mathbf{x}_n=\mathbf{x}_{\rm f}+\delta \mathbf{x}_n$ around the fixed point. The resulting variational equations read
\begin{align}\label{eq:vareqf}
    \delta \mathbf{x}_{n+1}=\mathbf{A} \delta \mathbf{x}_n \equiv \left.\frac{\partial \mathbf{F}}{\partial \mathbf{x}_n}\right|_{\mathbf{x}_n=\mathbf{x}_{\rm f}}\delta \mathbf{x}_n\,,
\end{align}
where the Jacobian matrix calculated at the fixed point and $\mathbf{A}$ is known as the {\it monodromy matrix}. The solution of this equation is found by projecting the initial conditions into the eigenbasis of $\mathbf{A}$ and the individual components then evolve as $\propto \lambda^n$, where $\lambda$ are the respective eigenvalues. In other words, the stability of an eigendirection depends on whether $|\lambda|$ is smaller or larger than one.

For Poincar{\'e} maps generated by Hamiltonian systems it holds that $\det(\mathbf{A})=1$. In the case of two-dimensional maps this implies that the eigenvalues will appear in pairs of the type\footnote{In the case of higher-dimensional Hamiltonian maps the eigenvalues have to come in quartets of the form $\lambda,1/\lambda,\lambda^*,1/\lambda^*$.} $\lambda_1 , \lambda_2 =  1/\lambda_1$. The explicit formula for the two eigenvalues reads
\begin{align}
    \lambda_{1,2}=\frac{\textrm{Tr}(\mathbf{A})\pm\sqrt{\textrm{Tr}(\mathbf{A})^2-4}}{2}\, ,
\end{align}
where $\textrm{Tr}(\mathbf{A})$ is the trace of the monodromy matrix.
\begin{itemize}
    \item If $|\textrm{Tr}(\mathbf{A})|<2$, then $\lambda_{1,2}\, \in \mathbb{C}$ and the eigenvalues can then be rewritten as $\lambda=\exp^{\pm i \vartheta}$, where $\vartheta=\cos^{-1}\left(\frac{\textrm{Tr}(\mathbf{A})}{2}\right)$ indicates the angular velocity with which the nearby points are rotating around the fixed point. Due to this rotation the point is sometimes called elliptic, but physically it corresponds to a {\it stable point} and the phase-space rotations around it correspond to small oscillations in the configuration space.
    \item If $|\textrm{Tr}(\mathbf{A})|=2$, then $\lambda_{1,2}=1$ or $\lambda_{1,2}=-1$. The point in this case is indifferently stable. Often this means the point is ``fixed by kinematic coincidence''. In other cases the appearance of an indifferently stable point means that variations of the system parameters will induce a topological transition in the phase-space flow.
    \item If $|\textrm{Tr}(\mathbf{A})|>2$, then $\lambda_{1,2}\, \in \mathbb{R}$. Each eigenvalue corresponds to an eigenvector defining an eigendirection. If $|\lambda_{1}|>1$, then $\lambda_{1}$ corresponds to an unstable eigendirection and $\lambda_{2}= 1/\lambda_1$ to a stable one. These eigendirections define a hyperbolic flow around the fixed point, therefore the point is called hyperbolic. However, physically the point corresponds to an unstable equilibrium, hence it is called an \ix{{\it unstable point}}; the presence of both the diverging and approaching directions typically correspond to the same process just with flipped directions of time.
\end{itemize}

\subsubsection{Stability of periodic trajectories and fixed points in continuous systems}

Let us consider a linear perturbation $\mathbf{y}=\mathbf{x}+$ on any point $\mathbf{x}$ of the phase space $\mathcal{S}$ of a Hamiltonian dynamical system. This perturbation lies in the tangent space $\mathcal{T}_\mathbf{x}\mathcal{S}$ to the phase space at the point $\mathbf{x}$. The \ix{{\it deviation vector}} $\pmb{\xi}$ can be evolved along the flow $\mathcal{F}_t$ by applying a linear operator from the tangent space at one point along the trajectory to a tangent space at a later point $\mathcal{D}_t:\, \mathcal{T}_\mathbf{x}\mathcal{S}\rightarrow \mathcal{T}_{\mathcal{F}_t(\mathbf{x})}\mathcal{S}$. The action of this operator takes the deviation vector $\pmb{\xi}(t_0)$ at time $t_0$ and evolves it to $\pmb{\xi}(t)=\mathcal{D}_t \pmb{\xi}(t_o)$ at time $t$. The respective evolution equations are given by a variation of eq.~\eqref{eq:HamCont} and read
\begin{align}
 \frac{\d \pmb{\xi}}{\d t}=\left. \frac{\partial \mathbf{f}}{\partial \mathbf{x}}\right|_{\mathbf{x}(t)}\pmb{\xi}\,.
\end{align}
The vector $\pmb{\xi}$ is then interpreted as the deviation between two infinitesimally close solutions of the system. In return, this deviation can show as how sensitive is a part of a dynamical system to its initial conditions. 

Let us first discuss the stability of a strictly fixed point $\mathbf{x}_{\rm f}$ such that $\d \mathbf{x}/\d t = \mathbf{f}(\mathbf{x}_{\rm f}) = 0$ where the evolution is simplified to
\begin{align}
 \frac{\d  \pmb{\xi}}{\d t}= \mathbf{B} \pmb{\xi} \equiv \left. \frac{\partial \mathbf{f}}{\partial \mathbf{x}}\right|_{\mathbf{x}_{\rm f}}\pmb{\xi}\, .
\end{align}
The fact that the matrix $\mathbf{B}$ is constant implies that the linearly independent solutions for $\pmb{\xi}$ are proportional to $e^{\kappa t}$, where $\kappa$ are eigenvalues of $\mathbf{B}$. Since $\mathbf{B}$ is generated from a Hamiltonian system \eqref{eq:HamCont}, its eigenvalues come in ``loxodromic quartets'' $\kappa, \kappa^*,-\kappa,-\kappa^*$ (degenerate doublets are also possible if either $\kappa \in \mathbb{R}$, or $\kappa \in \mathbb{I}$). If the the real part of $\kappa$ is positive or negative, the respective eigenvector corresponds to a stable or unstable direction respectively. 

Consider a fixed point of a Hamiltonian with a single degree of freedom, that is, a two-dimensional phase space. Now we have only three options similar to the fixed point in discrete systems:
\begin{itemize}
\item $\kappa_{1,2} \in \mathbb{I}, \,  \kappa_1 = \kappa_2* = -\kappa_2$. We define $\omega_{1,2} =  \kappa_{1,2}/i$ so that the solutions are rotating at angular velocity $\omega$ about the fixed point. Again this case corresponds to a stable equilibrium.  
\item $\kappa_1 = \kappa_2 = 0$. This is again an indifferently stable fixed point. 
 \item $\kappa_{1,2} \in \mathbb{R}$, $\kappa_1 = -\kappa_2$. The eigendirection corresponding to the positive $\kappa$ is unstable and the one to the negative stable, the flow is hyperbolic about $\vec{x}_{\rm f}$ and it corresponds to an unstable equilibrium.
\end{itemize}
An example of a stable fixed point is $x=0,p_x=0$ in Fig. \ref{fig:NLpendulum}. In the same figure an unstable fixed point can then be found at $x=\pi, p_x=0$. 

\begin{figure}
    \centering
    \begin{minipage}{\textwidth}
    \centering
    \raisebox{-0.42\height}{\includegraphics[width=0.47\textwidth]{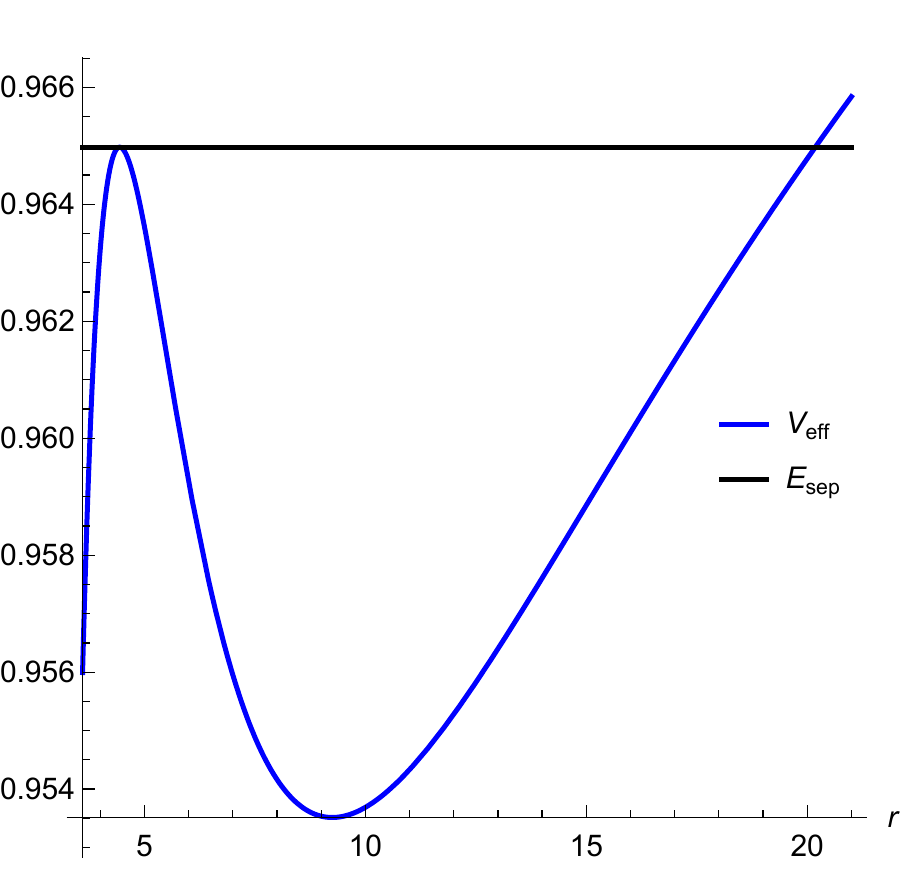}}
    \raisebox{-0.5\height}{\includegraphics[width=0.47\textwidth]{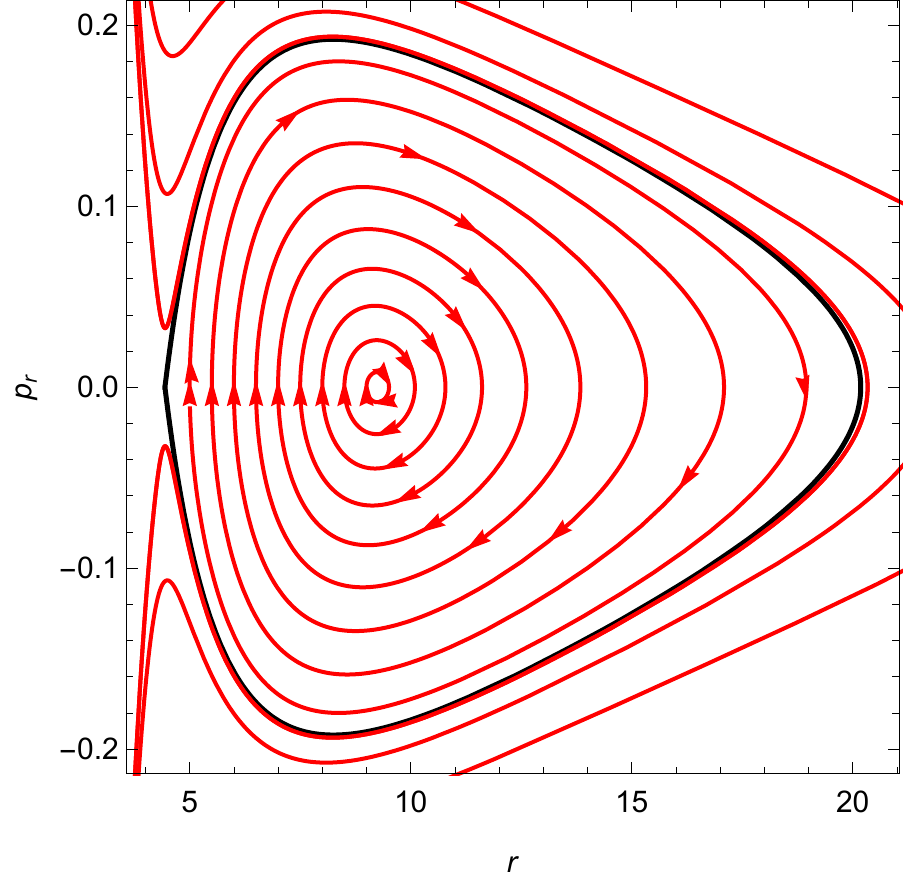}}
    \end{minipage}
    \caption{Left: The Schwarzschild effective potential for a free test particle at specific angular momentum $L=3.7M$ (blue) and the corresponding separatrix specific energy $E_{\rm sep}$. The potential is defined using specific energy $E$ as $V_{\rm eff} = E(p_r=p_z= 0,L=3.7 M)$ (see Eq.~\eqref{eq:EffPotA}). Right: The corresponding motion in the radial $r-p_r$ phase space. The eternally bound motion is separated from the plunging motion by the homoclinic separatrix (black), which originates at the unstable circular orbit at $r\approx 4.4M$. The stable circular orbit is at $r\approx 9.4M$ and it manifests as a stable fixed point in the diagram. Another separatrix between motion escaping and non-escaping to infinity would appear at $E=1$ outside of the plot range.}
    \label{fig:Schwarzport}
\end{figure}

Let us now turn to the stability of periodic orbits in continuous systems. There we can define a discrete map of the form \eqref{eq:maps} as the evolution of the whole system by the period of the orbit $T$, $\mathbf{F}(\mathbf{x}) = \mathcal{F}_{T}(\mathbf{x})$. The periodic orbit then reduces to a fixed point on this discrete map and we can easily see that its monodromy matrix $\mathbf{A}$ is equal to $\mathcal{D}_T$. Hence, up to a few cosmetic changes, the stability theory of periodic orbits is identical to that of fixed points of discrete maps as given in section \ref{sec:discstab}. Additionally, it often turns out that in separable systems periodic orbits can be examined in separable sub-spaces where they appear as fixed points. An important example of this would be circular geodesics in black hole space-times, in which the circular orbits appear as fixed points in the radial sector. The stability theory of periodic orbits in these separable sub-spaces is then identical to that of continuous-system fixed points discussed above. An example of such a reduction in Schwarzschild space-time is given in Fig. \ref{fig:Schwarzport}. 

\subsubsection{Stable and unstable manifolds}
We have discussed the linear stability of fixed points and periodic orbits in discrete and continuous dynamical systems. However, how does that relate to a more global, nonlinear picture?
According to the Hartman-Grobman theorem the \textit{qualitative} picture of the motion we obtain from the linearisation around fixed points and periodic orbits is always correct in some small neighborhood of the point \cite{Grobman59,Hartman60}. Furthermore the stable manifold theorem guarantees that one can prolong the stable and unstable directions into the so-called \ix{\textit{stable and unstable asymptotic manifolds}} \cite{Pesin77}. These are formally defined as follows:
\begin{itemize}
   \item The stable asymptotic manifold $\mathcal{M}^s$ is the set of points that asymptotically approach the unstable fixed point as $t\to \infty$.
   \item The unstable asymptotic manifold $\mathcal{M}^u$ is the set of points that approach the unstable fixed point as $t\to -\infty$.
\end{itemize}
It can be shown that the stable asymptotic manifold of one fixed point cannot cross itself or the stable asymptotic manifold of another fixed point. The same holds for the unstable asymptotic manifolds. However, \textit{stable and unstable} asymptotic manifolds can cross each other. Crossings of the same fixed point are then called \ix{\textit{homoclinic points}} and crossings of stable and unstable manifolds belonging to different fixed points are called \ix{\textit{heteroclinic points}} 

The character of the crossing is crucial. If $\mathcal{M}^u$ and $\mathcal{M}^s$ have a tangential intersection (their tangent manifolds coincide at the crossing), then they are necessarily just a part of a single smooth homoclinic or heteroclinic manifold and each of their points is part both of $\mathcal{M}^u$ and $\mathcal{M}^s$. These cases are known also as \ix{{\it separatrices}}. However, if they cross \textit{transversely} (at least parts of their tangent manifolds are independent at the crossing), then they both cannot form the same smooth manifold and there has to be an infinite number of such crossings. The result is the infamous \textit{homoclinic (or heteroclinic) tangle} where $\mathcal{M}^u$ and $\mathcal{M}^s$ are folded in an infinitely intricate manner into each other \cite{smale1965} giving rise to chaotic orbits. This is discussed in detail in section~\ref{sec:chalayr}.

\subsubsection{Stability of generic trajectories}

In some sense, it is also possible to measure the stability of {\em any orbit in phase space}. If the norm of the deviation vector $\pmb{\xi}$ introduced in the last section grows linearly with time, then the orbit is characterized as \ix{{\it regular}} (mildly sensitive to perturbations), while if it grows exponentially it is either \ix{{\it chaotic}} (highly sensitive to perturbations). A  chaotic orbit is an orbit that is not periodic while being highly sensitive to perturbation, otherwise it is simply an unstable periodic orbit. An indicator of chaos based on the orbit stability is the maximal Lyapunov Characteristic Exponent
\begin{align}
 \mathrm{mLCE}= \max_{\pmb{\xi}(t_0)}\lim_{t\to \infty} \frac{1}{t}\log{\frac{|\pmb{\xi}(t)|}{|\pmb{\xi}(t_0)|}}\, , \label{eq:mLCE}
\end{align}
where $|\pmb{\xi}|$ is some norm of the deviation vector (the result is independent of the choice of the norm). It is straightforward to see that $\mathrm{mLCE}\to 0$ for regular orbits, while for the unstable ones $\mathrm{mLCE}$ will converge to a constant value equal to the exponent of the exponential growth of $\pmb{\xi}$.   

In practice, $\mathrm{mLCE}$ for non-periodic orbits has to be evaluated numerically, and then the limit is approximated only by a finite-time integration. As a result, we can only detect characteristic exponents $\gtrsim 1/t_{\rm int}$, where $t_{\rm int}$ is the integration time. In other words, we can often only notice chaos if it is sufficiently strong and we may be unable to numerically distinguish between very mild instability and regularility. Also, even though the mLCE is defined as the maximum over the initial $\pmb{\xi}$, a generic initial condition will very often have a non-zero projection into the unstable part of the deviation subspace, which then always dominates the late-time growth. It is thus sufficient to compute the limit in \eqref{eq:mLCE} just for a handful of linearly independent vectors to determine the mLCE.

\subsection{KAM, Poincar\'{e}-Birkhoff theorem and chaos} \label{sec:chaos_th}

Chaos arises around resonances or unstable equilibria if an initially integrable system is perturbed. The basic features of the transition from integrability to non-integrability is dominated by two theorems: the \ix{{\it Kolmogorov-Arnold-Moser (KAM) theorem}}  \cite{Kolmogorov54,Moser62,Arnold63} and the \ix{{\it Poincar\'{e}-Birkhoff theorem}} \cite{Poincare12,Birkhoff13}.

\subsubsection{KAM theory and Birkhoff chains} \label{sec:KAMBirk}
Let us take an autonomous integrable Hamiltonian system $H_0(\mathbf{J})$ with $N$ degrees of freedom expressed in action-angle variables $\mathbf{J},\pmb{\theta}$ fulfilling the non-degeneracy conditions (Eq.~\eqref{eq:nondeg}). Now let us consider a close smooth Hamiltonian system of the form
\begin{align}\label{eq:HamPer}
    H(\pmb{\theta},~\mathbf{J})=H_0(\mathbf{J})+\epsilon H_1 (\pmb{\theta},~\mathbf{J})\, ,
\end{align}
where $\epsilon \ll 1$. The KAM theorem tells us that if the perturbation $\epsilon$ is sufficiently small, then there exists a $K(\epsilon)\lesssim \mathcal{O}(\sqrt{\epsilon})$ and a $d>N-1$ such that the set of tori satisfying the Diophantine condition
\begin{align}
    \left|\sum^N_{i=1} k_i \omega^i \right|>\frac{K(\epsilon)}{|k|^{d}}\,,
\end{align}
 will survive the perturbation with only small deformations. These tori are called KAM tori and their depiction on a Poincar\'{e} section KAM curves.  
In the KAM theorem the Diophantine condition ensures that the surviving tori are sufficiently far away from a resonance. However, the condition is fulfilled in a $\sim 1- \mathcal{O}(\sqrt{\epsilon})$ fraction of the volume of the phase space. Thus, the qualitative character of the motion in the system is mostly conserved under small perturbations.

Nevertheless, generally there is also an $\mathcal{O}(\sqrt{\epsilon})$ volume around resonances where the character of the motion changes qualitatively, and this change is described by the Poincar\'{e}-Birkhoff theorem \cite{Poincare12,Birkhoff13}. From the infinite number of orbits on a resonant torus only $2 n, n\in \mathbb{N}$ stay periodic, $n$ of which are stable, and $n$ of which unstable. In a subspace orthogonal to the periodic orbits, the phase space structure is reminiscent of a non-linear pendulum (see section \ref{susec:tools}).

This can be easily observed on Poincar{\'e} surfaces of section of systems of 2 degrees of freedom. Specifically, all unperturbed orbits on a resonance with $\omega^1:\omega^2 = r:s$ will have a periodicity $j$ equal to either $r$ or $s$ on the section, depending on its construction.  After the perturbation, only an even number $2nj$ of periodic points will survive on the section; half of them\footnote{Remember that every periodic orbit corresponds to $j$ points on the section. For instance, in the case $n=1$, all the stable and unstable points correspond to a single stable or unstable phase space trajectory respectively.} will be stable and the other half unstable. This chain of stable and unstable points in a resonance is called a Birkhoff chain.

\subsubsection{Chaotic layers}\label{sec:chalayr}

As discussed in section~\ref{sec:stab}, unstable points are anchoring points of stable and unstable asymptotic manifolds\footnote{This is true both for unstable trajectories in resonances as well as unstable periodic trajectories and fixed points in the integrable system before their perturbation.}, which is the birthplace of chaos for non-integrable systems. When and if the stable and unstable manifolds intersect transversely at one location in phase space, they necessarily have to do so an infinite number of times, which causes an infinitely folded fractal-like structure called a homoclinic (or heteroclinic) tangle. A homoclinic tangle is non-integrable and implies chaos. 

For example, black hole space-times naturally contain unstable circular orbits. Once these orbits are slightly pushed in the radial direction while keeping their energy and angular-momentum constant, they become either asymptotically approaching or diverging zoom-whirl orbits. For unstable circular orbits with specific energy below one the orbits that are pushed radially outwards will at first spiral out to a finite distance from the original orbit, but eventually return and start spiraling back in to the unstable circular orbits, which makes them {\em homoclinic}. These families of zoom-whirl orbits in phase space define homoclinic manifolds and it is here where we most often find chaos in perturbed black hole fields \citep[e.g.][]{semerak2010,witzany2015,polcar2019free}.

Nevertheless, for small perturbations, homoclinic chaos occurs only in a small layer around the asymptotic manifolds. In other words, the constants of motion $\mathbf{J}$ stay within $\mathcal{O}(\sqrt{\epsilon})$ bounds as compared to the non-integrable system and we mostly loose phase information (that is, error in $\pmb{\theta}$ quickly becomes $\mathcal{O}(1)$). However, as the perturbation further grows, KAM tori can dissolve into cantor sets called {\it cantori} \cite{Efthymiopoulos97}. The gaps in the cantori allow unstable and stable asymptotic manifolds of different resonances or unstable periodic orbits to cross each other and hence heteroclinic chaos appears. Heteroclinic chaos means that the chaotic trajectory can drift through the non-integrable layers to values of $\mathbf{J}$ that are $\gtrsim \mathcal{O}(\sqrt{\epsilon})$ far from the original values, which makes the unpredictability of the system more dire. It is also generally observed that the absolute value of Lyapunov exponents grows once heteroclinic chaos is established.  

There is no unanimously accepted definition of \ix{{chaos}}. For instance, in the discussion above we have informally used the definition of a chaotic orbit as one that is not periodic and has a positive Lyapunov characteristic exponent. On the other hand, we have also worked with the assumption that broken integrability in a volume of phase space of a Hamiltonian system implies chaoticity of all trajectories in that volume. Fortunately, the link between the two can be numerically established as follows:
\begin{enumerate}
    \item We observe the breaking of integrability as the trajectories not being bound to an $N$-dimensional torus in phase space. Instead, they densely fill a finite connected sub-manifold $\mathcal{X}$ of dimension larger than $N$. Such trajectories are known as transitive trajectories in $\mathcal{X}$ and their existence implies that $\mathcal{F}_t$ is {\em  transitive} in $\mathcal{X}$ \cite{silverman1992maps}. This means that for every two non-empty regions $\mathcal{U},\mathcal{V} \in \mathcal{X}$, there exists a $T$ such that $\mathcal{F}_T(\mathcal{U}) \cap \mathcal{V} \neq 0$.
    \item If we further assume that periodic orbits are dense in $\mathcal{X}$ (while being of zero measure), it follows that $\mathcal{F}_t$ has sensitive dependence on initial conditions in $\mathcal{X}$ \cite{Banks92}. Here sensitivity means  that  there  is  a distance $\delta$ such that in an arbitrarily close neighborhood of any $\mathbf{x}\in \mathcal{X}$ there always exists a point $\mathbf{y}$ and a constant $T$ such that $|\mathcal{F}_T(\mathbf{x}) - \mathcal{F}_T(\mathbf{y})|>\delta$, where $|\cdot|$ is some metric distance on $\mathcal{X}$.  
\end{enumerate}
Of course, the second assumption that the non-integrable region is densely filled with periodic orbits is non-trivial. A dense set of periodic points can only be proven to exist along the asymptotic manifolds when a homoclinic tangle occurs due to a transversal intersection of the stable and unstable manifolds \cite{smale1965}. It is plausible that this structure is promoted to the rest of the non-integrable volume, but not rigorously proven. However, there is ample numerical evidence that non-integrability always implies sensitive dependence on initial conditions \cite{contopoulos2004}.

\subsection{Tools to study resonances} \label{susec:tools}

\begin{figure} 
    \centering
    \includegraphics[width=0.49\textwidth]{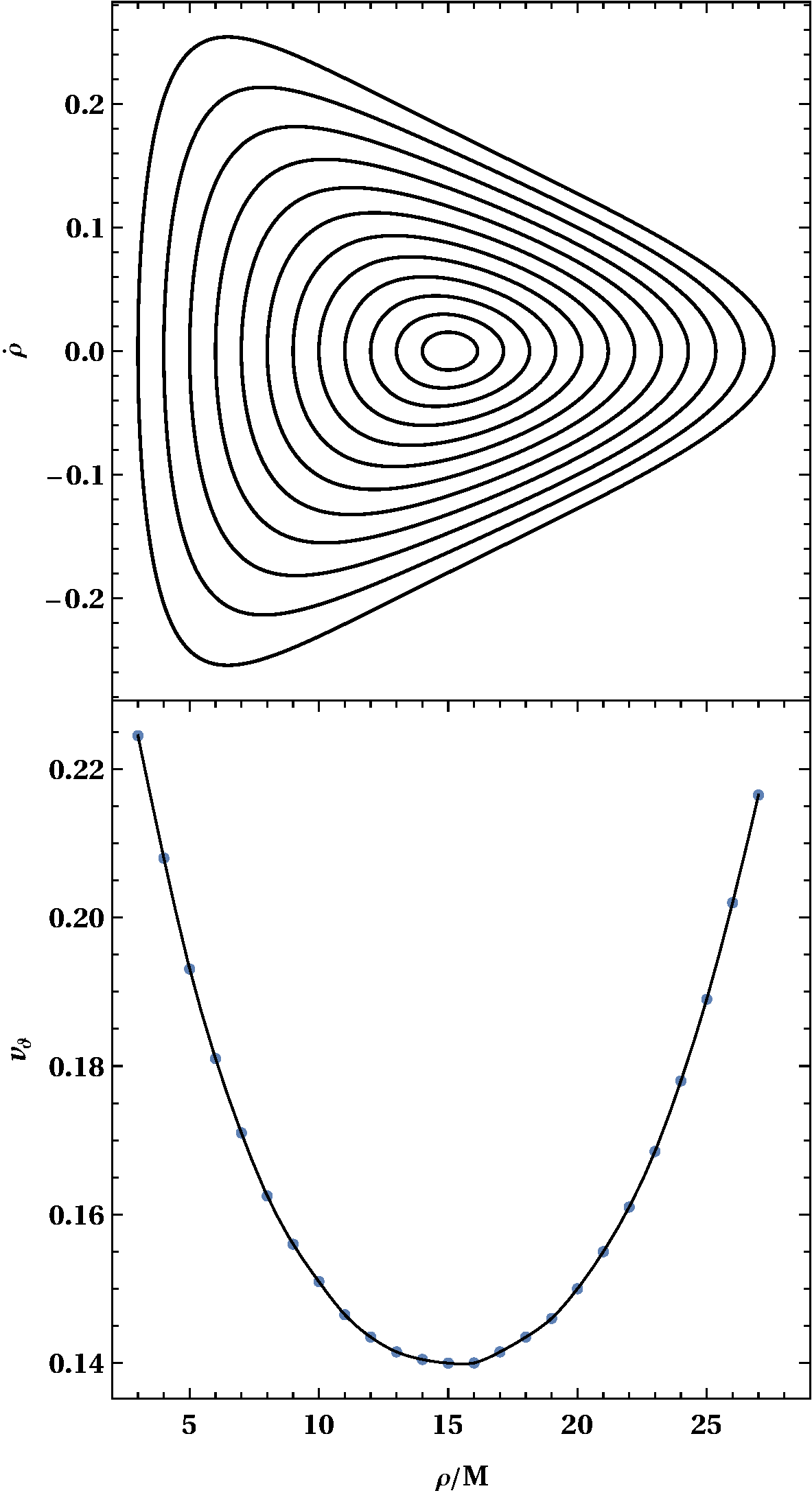}    
     \includegraphics[width=0.49\textwidth]{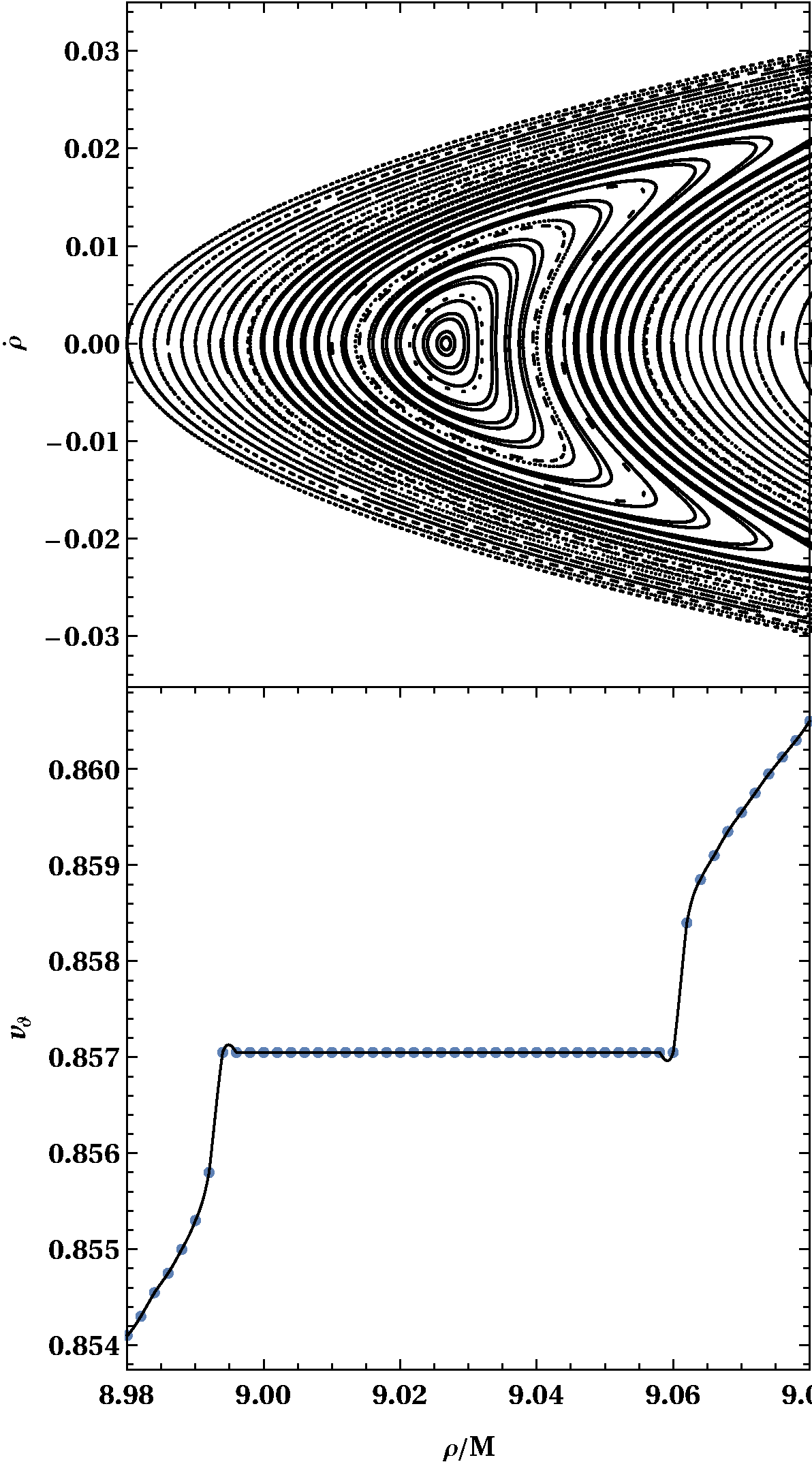}    
    \caption{The top panels show Poincar\'{e} sections, while the bottom panels show rotation curves calculated along the $\dot{\rho}=0$ line. All panels are calculated for a set of MSM parameters $M=1,~a=0.999~M$ and orbital parameters $L_z=3~M$, $E=0.97$. The left set of panels are plotted for $b=0$, while the right for $b=2.1~M$. The Poincar\'{e} section of the left case is dominate by KAM curves (top panel), thus when one starts from the centre of the foliation $(\rho\approx 15~M,~\dot{\rho}=0)$ and moves away along $\dot{\rho}=0$ the respective rotation curve appears to be strictly monotonic (bottom panel). The right case focuses on a resonance. The top panel is dominated by an island of stability of the $6/7$ resonance, which is reflected on the rotation curve in the bottom panel by the characteristic plateau.}
    \label{fig:sec_rot}
\end{figure}

Since the resonances are the places where chaotic motion arises, it would be useful to have a tool to spot them. A natural approach would be to survey the phase space by identifying fundamental frequencies of motion in the Fourier transform of the trajectories. However, the interpretation of the numerical results of this procedure can be tricky if the system is not in action-angle variables \cite{LGAC10}. In systems of two degrees of freedom, one can instead use a Poincar\'{e} surface of section to evaluate the so-called \ix{{\it rotation number}} $\nu_{\vartheta}$. The rotation number $\nu_\vartheta$ is defined as the ratio between the two fundamental frequencies of the system. According to this method one first identifies the center of a main island of stability, i.e. the fixed point $\mathbf{x}_{\rm c}$ on the Poincar\'{e} section around which the majority of the invariant curves are nested, and then finds the rotation angles\footnote{The angles can be defined with respect to any reasonable polar coordinates centered on $\mathbf{x}_c$.} $\vartheta_i := \mathrm{ang}\left[\left(\mathbf{x}_{i+1}-\mathbf{x}_{\rm c}\right), \left(\mathbf{x}_{i}-\mathbf{x}_{\rm c}\right)\right]$ between successive intersections $\mathbf{x}_i$ of the trajectory with the section with respect to $\mathbf{x}_{\rm c}$. The rotation number is then calculated as the average of the rotation angles
\begin{equation}\label{eq:angular_moment}
    \nu_\vartheta = \lim_{N\to\infty} \frac{1}{2\pi N}\sum_{i=1}^N \vartheta_i \:.
\end{equation}
For a non-degenerate foliation of an integrable system the rotation number changes strictly monotonically as one moves away from $\mathbf{x}_{\rm c}$. The dependence of $\nu_\vartheta$ on the distance from the center of the island is also known as the \ix{{\it rotation curve}} (see left set of panels in Fig.~\ref{fig:sec_rot}). Under perturbation the curve stays qualitatively similar to the unperturbed system up to resonances, near which new features appear. Namely, chaotic layers in the resonance appear as a random fluctuations of the rotation curve, while the islands of stability of a Birkhoff chain will create plateaus of constant rotation number values. The width of the plateau in weakly chaotic regions, when measured properly, corresponds quite accurately to the $\mathcal{O}(\sqrt{\epsilon})$ width of the resonance. 

The \ix{\textit{width of a resonance}} is a very useful quantity, since  in a perturbed Hamiltonian system~\eqref{eq:HamPer} the width of the resonances relates to the perturbation parameter $\epsilon$. In the following passage, we sketch the basic steps to reach this relation, for further details the interested reader is referred to \cite{Morbidelli02,Arnold06}.

Let us consider a 2 degrees of freedom Hamiltonian system, then the system~\eqref{eq:HamPer} reduces to 
\begin{align}\label{eq:HamPer2}
    H = H_0\left(J_1, J_2\right) + \epsilon H_{1}\left(J_1, J_2, \theta^1,\theta^2\right)\, .
\end{align}
Assuming that at the action values of the unperturbed Hamiltonian system $H_0$ $J_1=J_1^{\rm r},~J_2=J_2^{\rm r}$ lies a resonance $k_1 \omega^1+k_2 \omega^2=0$, we will rotate the action-angle variables as
\begin{align}\label{eq:TranRot}
    &\tilde{J}_1 = \frac{J_1}{2 k_1} +\frac{J_2}{2 k_2}\,,\; \tilde{\theta}^1 = k_1 \theta^1 + k_2 \theta^2 \,,\\
    & \tilde{J}_2 = \frac{J_1}{2 k_1} -\frac{J_2}{2 k_2}\,,\; \tilde{\theta}^2 = k_1 \theta^1 - k_2 \theta^2 \,.
\end{align}
One can easily see that these new variables are canonical by evaluating the symplectic form $\d\tilde{J}_1\wedge\d \tilde{\theta}^1 + \d\tilde{J}_2\wedge\d \tilde{\theta}^2$. It should also be noted that the $\theta$ coordinates have to be wound more than once to reach periodicity in $\tilde{\theta}$, which then means that $\tilde{\theta}^1$ and $\tilde{\theta}^2$ turn out to be $2(k_1^2 + k_2^2)\pi $ and $4 |k_1 k_2| \pi$-periodic respectively. 

In the new variables the resonance condition has been reduced to $\tilde{\omega}_1=0$. This implies that at the resonance for the perturbed system holds $\dot{\tilde{\theta}}_1 = \mathcal{O}(\epsilon)$. By applying an ``averaging'' near-identity transform $\tilde{J}_2 \to  \tilde{J}_2 + \mathcal{O}(\varepsilon),\, \tilde{\theta}^2 \to \tilde{\theta}^2 + \mathcal{O}(\varepsilon)$ we can eliminate the phase $\tilde{\theta}^2$ (see \cite{Arnold06} and section \ref{sec:resopass}). This averaging will render our system~\eqref{eq:HamPer2} to be approximated by the integrable system $H = H_0\left(\tilde{J}_1,\tilde{J}_2\right) + \epsilon H_{1}\left(\tilde{J}_1, \tilde{J}_2, \tilde{\theta}^1\right)$, in which the action $\tilde{J}_2$ is a constant of motion. 

The next step is to expand this Hamiltonian function in terms of the deviation of the action $\Delta\tilde{J}_1=\tilde{J}_1-{\tilde{J}_1}^{\rm r}$ from the resonance value to the leading order, which results in $H = \frac{\beta}{2}\left(\Delta \tilde{J}_1\right)^2 + \epsilon F\left(\tilde{\theta}^1\right)$, where $\beta,F$ depend on the constants $\tilde{J}_2,\,\tilde{J}_1^{\rm r}$. To arrive to the final form of the Hamiltonian, $F$ is expanded in a Fourier series from which we keep only the leading harmonic and introduce a phase shift to $\tilde{\theta}^1$ to obtain
\begin{equation}\label{eq:osc_phase1}
    H = \frac{\beta}{2}\left(\Delta \tilde{J}_1\right)^2 + \epsilon \alpha\cos\left(n\, \tilde{\theta}^1\right) \,,
\end{equation}
where $n$ corresponds to the one shown when discussing the Poincar\'{e}-Birkhoff theorem in section \ref{sec:KAMBirk}.
Without loss of generality, we can choose conventions such that $\alpha,\,\beta,\,\epsilon$ are positive parameters and the function~\eqref{eq:osc_phase1} is then essentially the Hamiltonian of a nonlinear pendulum. Hence, what we have shown that the phase portrait of a resonance can be approximately mapped to that of a pendulum. 

For $\Delta \tilde{J}_1=0$ the minima and the maxima of $\cos\left(n\tilde{\theta}^1\right)$ correspond to stable and unstable fixed points respectively. From the unstable fixed points stem separatrices, separating the near-resonant Birkhoff chain from the rest of the KAM tori. The location of the separatrices is approximately given as the level set $H=\epsilon \alpha$ of the Hamiltonian~\eqref{eq:osc_phase1}, which yields
\begin{align}\label{eq:separ}
    \left(\Delta \tilde{J}_1|_{\rm sep}\right)^2=\frac{2 \epsilon \alpha}{\beta}\left(1-\cos\left(n\, \tilde{\theta}^1\right)\right)
\end{align}
The width of the resonance is defined as the difference between the maximum and the minimum value of $\Delta \tilde{J}_1$ on the separatrix (i.e. when $\cos\left(n\, \tilde{\theta}^1\right)=-1$)
\begin{equation}\label{eq:pert_to_wid}
    \mathrm{width} := \max\left(\Delta \tilde{J}_1|_{\rm sep}\right) - \min\left(\Delta \tilde{J}_1|_{\rm sep} \right) = 4\sqrt{\frac{\alpha}{\beta}}\sqrt{\epsilon}  \,.
\end{equation}

Another way to find the width of the resonance is from the opening angle $\mathrm{d} \Delta \tilde{J}_1|_{\rm sep}/ \mathrm{d} \tilde{\vartheta}^1 $ between the separatrices at an unstable point. For example, for small deviations from the unstable point at $\tilde{\theta^1}=0$, Eq.~\eqref{eq:separ} reduces to 
\begin{align}
  \left(\Delta \tilde{J}_1|_{\rm sep, \tilde{\theta}^1 \to 0}\right)^2=\frac{\epsilon \alpha}{\beta}\left(n\, \tilde{\theta}^1\right)^2=\frac{\left(\mathrm{width}\, n\, \theta^1\right)^2}{4}  \, .
\end{align}
Hence, by having the coordinates $\Delta \tilde{J}_1,\, \theta^1$ of the separatrix near the unstable point we can estimate the width of the resonance.

Let us also establish a direct correspondence with the discussion of Birkhoff chains in section \ref{sec:KAMBirk}. If we go back to the initial system~\eqref{eq:HamPer2} and choose a Poincar\'{e} section on the plane\footnote{In the following, the labels $1,2$ can be swapped to obtain a different section.} $\theta^2=0$, then from the transformation~\eqref{eq:TranRot} we get $\tilde{\theta}^1=k_1\theta^1$ everywhere on the section. This implies that on the surface of section and at the $k_1 \omega^1+k_2 \omega^2=0$ resonance there will appear a total of $n  k_1$ islands of stability and $n k_1$ unstable points.

\section{Inspirals through resonances} \label{sec:resopass}
Here we briefly sketch the necessary theory and methods that are needed to efficiently integrate a set of dissipative equations through a resonance in a near-integrable system. At certain points we fast-forward to the discussion of the meaning of this theory in the context of gravitational-wave inspirals; the reader is welcome to skip these portions of the text and come to them later.

\subsection{A generic inspiral}
Consider an unperturbed (not necessarily Hamiltonian) dynamical system of $N$ degrees of freedom in action-angle coordinates $\mathbf{j}, \pmb{\theta}$. For clarity of certain complicated expressions in this section, we switch to components $j_a, \theta^b; a, b=1...N, \theta^b \in (0,2\pi]$ and we will use the Einstein summation convention. 

Now we subject this system to a generic non-Hamiltonian smooth perturbation of order $\epsilon \ll 1$ such that the equations of motion become
 \begin{align}
     & \dot{j}_a = \epsilon f^{(1)}_a(j_b, \theta^c) + \epsilon^2 f^{(2)}_a(j_b, \theta^c) + ...\,,\\
     & \dot{\theta}^b = \omega^b(j_a) + \epsilon g^b_{(1)}(j_a,\theta^c) + \epsilon^2 g^b_{(2)}(j_a,\theta^c)+...
 \end{align}
 In the language of the gravitational self-force, $\epsilon$ corresponds to the mass ratio, $f^{(n)}(j,\theta)$ then involves both the averaged and oscillating dissipative self-force \textit{and} some of the conservative self-force of order $n$. On the other hand, $g_{(n)}(j,\theta)$ involves only the oscillating dissipative and conservative parts of the self-force of order $n$. Now we are interested in the approximate evolution of this system over a long time $t_{\rm insp}$ that can be characterized as $t_{\rm insp} \sim 1/(\epsilon \omega)$. Specifically, we would like the error of the final phase $\theta^b(t_{\rm insp})$ to go to zero in the $\epsilon \to 0$ limit. In the completely generic case, one needs to switch between the treatment for non-resonant (weakly resonant) and strongly resonant parts of phase space. 
 
Recall that generic resonances are hypersurfaces in the action space characterized by Eq.~\eqref{eq:res_cond}, which in the current discussion reads
 \begin{align*}
     k_a \omega^a(j_b) = 0,\, k_a \in \mathbb{Z}^N\,.
 \end{align*}
Such hypersurfaces create a dense net in the phase space and in practice one needs to identify a finite set of ``strong'' resonances for a separate treatment. This is done by expanding the functions $f^{(n)}_a, g^b_{(n)}$ into trigonometric polynomials (a Fourier-coefficient expansion) over $\theta^b$. The magnitude of the terms in this expansion will quickly fall off due to the smoothness of the functions $f,g$ and one can thus split the expansion into a finite number of dominant $\mathcal{O}(1)$ terms and sub-dominant terms of relative order $\epsilon$. This can be viewed as a smoothing of the functions $f^{(n)}_a, g^b_{(n)}$ on the torus, where oscillations in the function values of relative magnitude $\lesssim \epsilon$ are pushed to higher order. Even though this discarding scheme is dependent on the absolute value of $\epsilon$, it is the only way to isolate a finite number of resonances and build an effective $\epsilon\to0$ limit of the perturbed system.\footnote{In fact, the understanding of this ``moving target'' character of the perturbation theory as $\epsilon \to 0$ is the one of the essential points of the proof of the famous Kolmogorov-Arnol'd-Moser theorem.} A \ix{strong resonance} at $\mathcal{O}(\epsilon^n)$ is then a resonance with a wavenumber $k_c$ which has a trigonometric polynomial of order $k_c$ in the dominant part of $f^{(n)}_a$ (resonances caused by $g^b_{(n)}$ cause minor resonances in our inspiral-type scenario). 

\subsection{\ix{Non-resonant motion}}
Away from strong resonances we can apply a near-identity transform $J_a = j_a + \epsilon \xi^{(1)}_a (j_b,\theta^c) + \epsilon^2 \xi^{(2)}_a (j_b,\theta^c),\, \Theta^b = \theta^b + \epsilon \eta_{(1)}^b (j_c,\theta^a) + \epsilon^2 \eta_{(2)}^b (j_c,\theta^a)$ such that the equations of motion attain the form \citep{Arnold06}
 \begin{align}
     & \dot{J}_a = \epsilon F^{(1)}_a(J_b) + \epsilon^2 F^{(2)}_a(J_b)  +\mathcal{O}(\epsilon^3)\,, \label{eq:dotJ}
     \\
     & \dot{\Theta}^b = \Omega^b(J_a) + \epsilon G^b_{(1)}(J_a) + \epsilon^2 G^b_{(2)}(J_a) +  \mathcal{O}(\epsilon^3) \,, \label{eq:dotPsi}
 \end{align}
 where $F,G$ are given as sums of averages of $f,g$ over the angles $\theta^b$ and various transformation terms. $\Omega^b$ can be chosen to be functionally identical with $\omega^b$ apart from the fact that $J_b$ was inserted instead of $j_b$ as the argument. The transformation vectors $\xi,\eta$ can then be computed from the requirement that the transformation leads to equations of motion of the form \eqref{eq:dotJ} and \eqref{eq:dotPsi}, which yields
 \begin{align}
     & \xi_a^{(1)} = \left\{f^{(1)}_a\right\}^\theta \,,
     \\
     & \eta^a_{(1)} = \left\{g_{(1)}^a + \frac{\partial \omega^a}{\partial j^b} \xi_b^{(1)} \right\}^\theta \,,
     \\
     & \xi_a^{(2)} = \left\{f^{(2)}_a + \frac{\partial f^{(1)}_a}{\partial \theta^b} \eta^b_{(1)} + \frac{\partial f^{(1)}_a}{\partial j_b} \xi_b^{(1)} - \frac{\partial \xi_a^{(1)}}{\partial j_b} F_b^{(1)} - \frac{\partial \xi_a^{(1)}}{\partial \theta^b} G^b_{(1)} \right\}^\theta \,,
     \\
     & \eta^a_{(1)} = \left\{g^a_{(2)} + \frac{\partial \omega^a}{\partial j_b } \xi_b^{(2)} + \frac{1}{2} \frac{\partial^2 \omega^a}{\partial j_b \partial j_c } \xi_b^{(1)} \xi_c^{(1)} - \frac{\partial \eta^a_{(1)}}{\partial j_b} F_b^{(1)} - \frac{\partial \eta^a_{(1)}}{\partial \theta^b} G^b_{(1)}   \right\}^\theta \,,
     \\
     & \left\{h(j_a,\theta^b)\right\}^\theta \equiv \sum_{k_b \neq 0} \frac{h_{k_b}(j^a)}{i k_b \omega^b(j^a)} e^{i k_b \theta^b}\,,
 \end{align}
 where $h_{k_b}$ are the Fourier coefficients of $h$ over $\theta^b$. The transformation is determined uniquely only up to ``integration constants'', and one can add arbitrary bounded smooth functions of $j_a$ (but not of $\theta^b$) to any of the $\xi,\eta$.
 
 The advantage of the transformed system \eqref{eq:dotJ} and \eqref{eq:dotPsi} is the possibility to integrate the new actions $J_a$ separately from the angles $\Theta^b$. The actions evolve on the inspiral time-scale $\sim J_0/(\epsilon F^{(1)}) \sim 1/(\epsilon \Omega)$ and their decay is usually stiff. On the other hand, the phases evolve on a time-scale $1/\Omega$ and their evolution has the character of a steady increase with a slowly changing slope. It is then obvious that the $J$-$\Theta$ split allows for an efficient choice of time steps and integration methods for each of the sub-problems.
 
 In fact, once the functional forms of $F_a^{(1)},F_a^{(2)},G^b_{(1)}$ are explicitly known, we can drop the $G^b_{(2)}$ terms, integrate the equations, and the resulting solutions $J_a(t),\Theta^b(t)$ will be globally $\epsilon$-close to the exact solution $j_a(t), \theta^b(t)$ over a time interval of order $1/\epsilon$ if no strong resonances are encountered \cite{Kevorkian12}. Even more, since none of the needed quantities depend on $\xi^{(2)}_a,\eta^b_{(2)}$, the second-order part of the transform does not need to be known explicitly for such a solution. In the language of gravitational self-force, this is equivalent to the statement that for accurate inspirals not passing through resonances one requires the full phase dependence of the first-order dissipative and conservative self-force, but only the average dissipative piece of the second-order self-force.

\subsection{\ix{Near-resonant motion}} 
It can be easily checked that the transformation functions $\xi^{(n)},\eta_{(n)}$ contain potentially singular terms. The leading-order singularity for $\xi^{(n)}$ is $\sim \epsilon^n f^{(1)}_{k}/(k_b \omega^b)^{2n-1}$ and for $\eta_{(n)}$ even $\sim \epsilon^n f^{(1)}_{k}/(k_b \omega^b)^{2n}$. When we approach sufficiently close to a strong resonance such that $k_a\omega^a\to 0$ for some $k_a$, the near-identity transform becomes ill-convergent and even completely meaningless when $k_a \omega^a \sim \sqrt{\epsilon}$. One thus needs to switch to a different description at some well-chosen point before the break-down. To ``hand over'' the original variables $j,\theta$ with sufficient accuracy, one should compute as many terms of the second-order transform $\xi^{(2)},\eta_{(2)}$ as possible without the knowledge of $g_{(2)}(j,\theta), f^{(2)}(j,\theta)$, since the known terms are also the most singular near resonance. Then if we choose a cut-off index $\beta\in (0,1/2)$ such that we cut-off the evolution at $k_a \omega^a \sim \epsilon^\beta$, we will hand over the phases $\theta$ with error terms of order $\sim \epsilon^{3-6\beta}$ and the actions with errors of order $\sim \epsilon^{3-5\beta}$. The optimal choice of $\beta$ will be discussed in section \ref{sec:errbud}.
 
Let us define the \ix{{\em resonant phase}} $\gamma \equiv k_a \theta^a$ and a set of non-resonant phases $\tilde{\theta}^{\tilde{a}},\,\tilde{a}=1...N-1$ obtained through various linear combinations of the original phases $\theta^a$ such that $\theta^a \mapsto \gamma,\tilde{\theta}^{\tilde{a}}$ is an invertible coordinate transform. We can then carry out a near-identity transform $j_a,\tilde{\theta}^{\tilde{b}},\gamma \mapsto \tilde{J}_a,\tilde{\Theta}^{\tilde{b}}, \Gamma$ eliminating the phases $\tilde{\theta}^{\tilde{a}}$ (or $\tilde{\Theta}^{\tilde{a}}$) analogous to the one for the full set $\theta^a$ above \citep{Arnold06}. However, in this case we will have no convergence issues near the resonance, since the second-order transformation terms have only denominators of the type $\tilde{k}_{\tilde{a}} \tilde{\omega}^{\tilde{a}},\, \tilde{k}_{\tilde{a}} \in \mathbb{Z}^{N-1}$ where $\tilde{\omega}^{\tilde{a}}$ are the non-resonant frequencies. We then obtain evolution equations of the form
\begin{align}
     & \dot{\tilde{J}}_a = \epsilon \tilde{F}^{(1)}_a(\tilde{J}_b,\Gamma) + \epsilon^2 \tilde{F}^{(2)}_a(\tilde{J}_b,\Gamma)  +\mathcal{O}(\epsilon^3)\,, \label{eq:dottildeJ}
     \\
     & \dot{\tilde{\Theta}}^{\tilde{b}} = \Omega^{\tilde{b}}(\tilde{J}_a) + \epsilon \tilde{G}^{\tilde{b}}_{(1)}(\tilde{J}_a,\Gamma) + \epsilon^2 \tilde{G}^{\tilde{b}}_{(2)}(\tilde{J}_a,\Gamma) +  \mathcal{O}(\epsilon^3) \,, \label{eq:dottildePsi}
     \\
     & \dot{\Gamma} = k_a \omega^a(\tilde{J}_a) + \epsilon \chi_{(1)}(\tilde{J}_a,\Gamma) + \epsilon^2 \chi_{(2)}(\tilde{J}_a,\Gamma)  + \mathcal{O}(\epsilon^3)\,, \label{eq:dotGamma}
 \end{align}
 where the meaning of the functions $\tilde{F},\tilde{G},\tilde{\Omega}$ is analogous to $F,G,\Omega$ in the previous paragraphs. Additionally, we see that the ``frequency'' of $\Gamma$-evolution is $k_a \omega^a \lesssim \mathcal{O}(\sqrt{\epsilon})$ in the resonant region, and that $\chi_{(1,2)}$ collects transformed terms of order $\epsilon,\epsilon^2$ respectively. This leads to $\Gamma$ often being called a ``semi-fast'' variable, since it evolves much slower than the regular phases $\tilde{\Theta}$ in the resonant region, but generally faster than $\tilde{J}_a$. The main advantage of this modified coordinate transform is that even though we increase the number of variables that need to be solved in the first step by one to the set $\Gamma,\tilde{J}_a$, at least the $N-1$ phases $\tilde{\Theta}^{\tilde{a}}$ can be still solved later in a separate step. Hence, we will now only focus on the solution for $\Gamma,\tilde{J}_a$.
 
 In the context of the gravitational self-force, we may not able to easily evaluate terms such as $\tilde{F}^{(2)}(\tilde{J},\tilde{\Gamma})$ everywhere, so we need to analyse the costs of omitting a part of them over the period of integrating through the non-resonant region. Specifically, it will be possible to evaluate $\tilde{F}^{(2)}(\tilde{J},\tilde{\Gamma})$ {\em exactly} when $k_a\omega^a = 0$, since then it amounts only to an infinite-time average over individual resonant trajectories, but we assume it will not be possible to evaluate this term anywhere else.\footnote{This assumption is not set in stone, it is in principle possible to evaluate derivatives of $\tilde{F}^{(2)}(\tilde{J},\tilde{\Gamma})$ by computing black hole perturbations based on a ``blurred'' stationary trajectory with nonzero $\dot{\tilde{J}},\dot{\Gamma}$, the same way $F^{(2)}(J)$ has to be computed on a blurred stationary trajectory with a non-zero $\dot{J}$ \cite{Miller20}.}

 \subsection{Error budget} \label{sec:errbud}
 Let us assume that we switch to the coordinates $\tilde{J},\Gamma, \tilde{\Theta}$, integrate through the resonance for a period $\Delta t_{\rm pass}$ and then switch back to the coordinates $J,\Theta$. Considering that the switch happens whenever $k_a \omega^a \sim \epsilon^\beta$ and the frequencies drift with a rate $\sim \epsilon \Omega_{,J} F$, we can estimate that $\Delta t_{\rm pass} \sim  \epsilon^{\beta - 1}$. Let us now estimate the errors by Taylor-expanding the solutions for $\tilde{J}(t),\Gamma(t)$ around the instant $t_{\rm r}$ at which $k_a \omega^a = 0$ exactly\footnote{For the actual evolution, it is more practical to numerically integrate the equations of motion, but the Taylor expansion provides a good way to estimate the overall error budget.}:
 \begin{align}
     & \tilde{J}_a(t_r + \Delta t_{\rm pass}) = \sum_{l=0}^\infty \frac{\tilde{J}^{(l)}(t_{\rm r})}{l!} (\Delta t_{\rm pass})^l \sim \sum_{l=0}^\infty \tilde{J}^{(l)}(t_{\rm r}) \epsilon^{l(\beta-1)} \,, \\
     & \Gamma_a(t_r + \Delta t_{\rm pass}) = \sum_{l=0}^\infty \frac{\Gamma^{(l)}(t_{\rm r})}{l!} (\Delta t_{\rm pass})^l \sim \sum_{l=0}^\infty \Gamma^{(l)}(t_{\rm r}) \epsilon^{l(\beta-1)} \,.
 \end{align}
We can then evaluate all the time-derivatives $\tilde{J}^{(l)},\Gamma^{(l)}$ from the equations of motion to a certain accuracy given that we are able to evaluate $\tilde{F}^{(1)}(\tilde{J},\Gamma), \chi_{(1)}(J,\Gamma)$ and all its derivatives. As mentioned above, we can also assume that we can evaluate $F^{(2)}(J,\Gamma)|_{t_{\rm r}}$ but none of its derivatives, which leads to a phase error term that is bounded by $\sim \epsilon^{3 \beta}$ and an action error term bounded by $\epsilon^{1+2\beta}$. Another source of error comes from the inability to evaluate $\tilde{G}_{(2)}(\tilde{J},\Gamma)$, which leads to a phase error of order $\epsilon^{1+\beta}$ and action error again of order $\epsilon^{1+2\beta}$. Additionally, one must also consider that the phase $\Gamma$ and actions $\tilde{J}$ were handed over with $\sim \epsilon^{3-5\beta}, \epsilon^{3-6\beta}$ errors respectively; this induces errors of order $\sim \epsilon^{2-4\beta}$ in both phases and actions after the resonant evolution. 

What is then the optimal value for $\beta$ given that we are interested in the overall \ix{inspiral faithfulness}? A smaller $\beta$ means that we integrate in the near-resonant coordinates for a longer time while possibly not increasing the accuracy of the total computation any more. On the other hand, a larger $\beta$ means a larger error is accumulated as one approaches and drifts away from the resonant region in the fully averaged coordinates. We thus need to find a value of $\beta$ such that the hand-over happens exactly at the point when further near-resonant integration would be redundant. However, the final answer also depends on whether we care more about the error in the phase or in the actions. 

Let us now assume that the inspiral encounters only a single strong resonance at a generic point, that is, a point such that there is still a $\sim 1/\epsilon$ time left after resonance exit. Then any error in the action upon resonance exit translates into the final inspiral phase with a factor $\sim 1/\epsilon$ while the resonant phase error is, at leading order, simply added to final phase error. It is then easy to see that the error budget is dominated by the actions, and it is optimized exactly when $\beta = 1/4$. This is because at that point the error of the near-resonant integration and the hand-over in the actions are both $\sim \epsilon^{3/2}$. In summary, assuming one has the complete first-order pieces of the perturbation and time-averaged dissipative pieces of the second-order perturbation, it is possible to squeeze the total inspiral phase error of the passage through a single resonance to $\sim \sqrt{\epsilon}$ in a well-defined procedure. However, ignoring the resonant terms entirely would lead to a $\sim 1/\sqrt{\epsilon}$ error in the inspiral phase.

An informed reader will notice that this scenario corresponds to something which is known as a \ix{\textit{transient resonance}}, that is, the resonant behaviour is dominant for a time $\sim 1/\sqrt{\epsilon}$. However, there may occur cases such that $k_a \omega^a$ remains $\lesssim \mathcal{O}(\sqrt{\epsilon})$ for a longer time. The time derivative of the resonant condition is
\begin{align}
    \frac{\mathrm{d}}{\mathrm{d} t} (k_a \omega^a(j_b))  = k_a \dot{\omega}^a  = \epsilon k_a  \frac{\partial \omega^a}{\partial j_b}  f_b(j_c, \theta) + \mathcal{O}(\epsilon^2)\,.
\end{align}
The assumption of our analysis then is that the average of $k_a  \dot{\omega}^a $ over the non-resonant phases is never below order $\epsilon$ everywhere where the resonant condition $k_a \omega^a = 0$ is met. If, however, a resonance has $k_a  \dot{\omega}^a = \mathcal{O}(\epsilon^2) $ anywhere, then it is in principle possible to obtain a \ix{\textit{sustained resonance}} such that the dominant behaviour is dominant for a time of the order of the \textit{entire} inspiral $\sim 1/\epsilon$ \cite{maarten2014}. We expect these to be non-generic cases that are negligible in realistic systems. However, if some symmetry makes this degeneracy of the equations of motion generic, the most efficient system for the integration over the time of the entire inspiral is simply the ``near-resonant'' system \eqref{eq:dottildeJ}-\eqref{eq:dotGamma}.

\subsection{Additional perturbations} \label{sec:addpet} 
Let us now consider additional perturbations of magnitude $\kappa \ll 1$ such that the equations of motion become
\begin{align}
     & \dot{j}_a = \epsilon f^{(1)}_a(j_b, \theta^c) + \epsilon^2 f^{(2)}_a(j_b, \theta^c) + \kappa h(j_b,\theta^c) +...\,,\\
     & \dot{\theta}^b = \omega^b(j_a) + \epsilon g^b_{(1)}(j_a,\theta^c) + \epsilon^2 g^b_{(2)}(j_a,\theta^c)+ \kappa l(j_a,\theta^c)+...
 \end{align}
We will classify the additional perturbation as conservative when $h(j_b,\theta^c)$ has a zero average over $\theta^c$ and thus the $\kappa$-terms do not contribute to the long-term decay of actions $j_b$, and as dissipative otherwise. Depending on the context, one can also have various $\sim\kappa \epsilon$ cross-terms as well. For instance, for inspirals in stationary axisymmetric space-times that are some deformations of Kerr space-time (see Section \ref{sec:deviating}) the perturbation is conservative and one can generally have $\kappa \gtrsim \epsilon$ and $\kappa\epsilon$ cross-terms would correspond to the fact that the self-force deviates from the Kerr self-force in the modified space-time. On the other hand, for modifications of gravity that only alter the radiation-reaction dynamics one would generally consider $\kappa \ll \epsilon$. It is obvious that away from resonances one can perform consecutive near-identity transforms that gradually eliminate the phase dependence of all the $\kappa$ and $\epsilon$ terms from the equations of motion and put them into an ``averaged'' form. Once again, however, separate treatment is required near resonances. 

Let us discuss only the effect of large conservative perturbations near resonances in more detail. A conservative perturbation with $\kappa\gtrsim \epsilon$ requires the switch to a near-resonant description of the equations of motion already when $k_a \omega^a \sim \kappa^{\beta'}, \beta'\in (0,1/2)$ and then the near-resonant integration has to be carried out over an interval $\Delta t_{\rm pass} \sim \kappa^{\beta'}/\epsilon$. In other words, the near-resonant behavior dominates the motion for an order of at least $\sim \sqrt{\kappa}/\epsilon$ cycles.  By using the techniques sketched above, it is then easy to show that the leading-order contribution of the $\kappa$-terms to the final inspiral phase due to the resonance passage are of order $\kappa^{1+\beta'}/\epsilon^2$. If the contribution of $\kappa$-terms at resonance are completely ignored, the system will have a phase error of order $\kappa^{3/2}/\epsilon^2$ at the end of the inspiral. To which degree this error can be removed depends on a more delicate analysis of the relative magnitudes of the $\epsilon$ and $\kappa$ terms and our ability to evaluate them.$\!$\footnote{Restricting now to the discussion of EMRIs and the gravitational self-force in non-Kerr space-times, one of the main issues would be the fact that there would be various $\sim \kappa^n \epsilon^l$ cross-terms due to the fact that one has to use different self-force than in the Kerr space-time. The dominant source of irremovable error would probably be an unknown $\sim \kappa \epsilon$ term in $\dot{j}$ correcting the radiation-reaction.}

\section{Orbital motion in Kerr spacetimes and perturbations}\label{sec:OrbKerr}

The current consensus is that the spacetime around a black hole is described by the Kerr solution \cite{Kerr63}. This paradigm, broadly used in the fields of astrophysics and gravitational-wave theory, is known also as the \ix{"Kerr black hole hypothesis"} \cite{Bambi11}.
The metric elements of the \ix{Kerr spacetime} in Boyer-Lindquist coordinates $(t,r,\vartheta,\varphi)$ read 
\begin{align}
   & g_{tt} =-1+\frac{2 M r}{\Sigma}\,,\:\: 
   g_{t\varphi} = -\frac{2 a M r \sin^2\!{\vartheta}}{\Sigma}\,,\:\: 
   g_{rr} = \frac{\Sigma}{\Delta}\,, \nonumber \\
   & g_{\varphi\varphi} = \frac{\Lambda \sin^2\!{\vartheta}}{\Sigma} \,,\:\: 
   g_{\vartheta\vartheta} = \Sigma\,,
    \label{eq:KerrMetric}
\end{align}
where
\begin{align}
  &\Sigma = r^2+ a^2 \cos^2\!{\vartheta}\,,\:\: 
  \Delta = \varpi^2-2 M r \,, \nonumber \\
  &\varpi^2 = r^2+a^2 \,,\:\: 
  \Lambda =\varpi^4-a^2\Delta \sin^2\!{\vartheta} \, ,  \label{eq:Kerrfunc}
\end{align}
$M$ is the mass and $a$ is the angular momentum per mass. The Kerr metric describes an asymptotically flat vacuum spacetime that is stationary, axisymmetric, and symmetric with respect to reflections about the equatorial plane $(\vartheta=\pi/2)$. The Kerr metric describes the field of an isolated rotating black hole as long as a horizon is covering the ring singularity, which holds for  $a<M$. For $a>M$ the Kerr spacetime corresponds to a naked singularity. For $a=0$ the \ix{Schwarzschild solution} is recovered and the Boyer-Lindquist coordinates are reduced to Schwarzschild coordinates.

A Hamiltonian function giving the geodesic motion of a massive test particle with respect to the proper time $\tau$ in curved spacetime reads
\begin{align}\label{eq:HamGeo}
    H=\frac{1}{2}g^{\nu \kappa}p_\nu p_\kappa=-\frac{1}{2}\mu^2\,,
\end{align}
where $\mu$ is the mass of the test particle. In the framework of General Relativity the system has four degrees of freedom. In the case of Kerr spacetime, the system is integrable, since there are four independent integrals of motion in involution. Namely, the stationarity and the axisymmetry imply that energy $E=-p_t=$ and angular momentum $L_z=p_\varphi$ along the symmetry axis $z$ are constants of motion. Additionally, the Hamiltonian itself expresses the conservation of the test particle's mass and the fourth constant 
\begin{align}
%    Q &=\mathcal{K}-(L_z-a E)^2\qquad \text{with} \nonumber \\
    \mathcal{K} &={p_\vartheta}^2+ \left(a E \sin{\vartheta}-\frac{L_z}{\sin{\vartheta}}\right)^2+a^2 \mu^2 \cos^2\!{\vartheta}  \nonumber \\
    ~ &= 2 \left(\varpi^2 E-a \, L_z\right] p_r- (\Delta {p_r}^2+\mu^2 r^2) 
\end{align}
discovered by Carter \cite{Carter68} reflects a hidden symmetry. In the non-spinning limit of Kerr spacetime, i.e. the spherically symmetric Schwarzschild spacetime, the \ix{Carter constant} reduces to the total angular momentum, which is constant as well. The existence of the four integral of motion suggests the system should be separable, i.e. we should be able to evolve each degree of freedom independently, but in order to achieve this the Carter-Mino time has to be employed \cite{Carter68,Mino03}. 

\subsection{Deviating Spacetimes} \label{sec:deviating}

The Carter constant is a unique feature of the Kerr spacetime, since it appears that there is no other stationary, axisymmetric and asymptotically flat spacetime in General Relativity that possesses similar "hidden" symmetry \cite{Carter68b,Frolov17}. Attempts to construct solutions possessing Carter constant by perturbing Kerr metric led to solutions obeying alternative theories of gravity, but not the Einstein's field equations \cite{Vigeland11,Johannsen13}. Even on the level of a Newtonian and electromagnetic analogue of Kerr, the Carter constant appears to be a unique feature of the Kerr-like Newtonian and electromagnetic fields \cite{Markakis14,Eleni20,lynden2000carter}. This implies that in the framework of General Relativity any deviation from Kerr spacetime destroys the integrability of the geodesic motion. 

\subsubsection{Bumpy black holes}
One way to parametrize a solution of Einstein's field equations that deviates continuously from the Kerr one is to introduce one or more parameters changing the Geroch-Hansen \ix{multipole moments of the Kerr field} \citep{hansen1974}
\begin{align}\label{eq:KerrMulExp}
 M_n+i S_n= M (i a)^n \,,\: n \in \mathbb{N} \, ,
\end{align}
where $M_n$ and $S_n$ are the mass and the current-mass multipole moments respectively. For example, the Manko-Novikov solution \cite{Manko92} introduces an extra parameter for each mass multipole moment. The fact that the solution is characterized by another parameter that can be seen in its structure outside the horizon necessarily implies that such solutions have at least broken horizons, otherwise the no-hair theorem would be violated. In the case of the Manko-Novikov solution, this manifests as a ring singularity on the horizon, which disappears as the extra multipole moments are switched off and the Kerr solution is recovered. Such solutions are often called \ix{bumpy or non-Kerr black holes} and reflect the possibility of having compact objects in the General Relativity framework that challenge the Kerr black hole hypothesis, even though it is currently unclear how such objects should form. 

As an example of a bumpy black hole, we are going to use a reduced version of an exact solution known as the \ix{Manko, Sanabria-G\'{o}mez, Manko (MSM) solution} \cite{MSM}. The original MSM spacetime depends on five real parameters: the mass $M$, the spin $a$ (per unit mass $M$), the charge, the magnetic dipole moment and the mass-quadrupole moment ${\cal Q}$. However, in the reduced version the charge and the magnetic dipole are set to zero. This allows the mass-quadrupole moment 
\begin{align} \label{eq:MassQuadRed}
  {\cal Q}=-M \left(\frac{(M^2-(a-b)^2)^2+4~M^2~b^2}{4(M^2-(a-b)^2)}-a~b+a^2\right) 
\end{align} 
to deviate from the Kerr one ${\cal Q}_\textrm{Kerr}=-a M^2$ by one free parameter $b$. The Kerr mass-quadrupole is retrieved from Eq.~\eqref{eq:MassQuadRed} for $b^2=a^2-M^2$.\footnote{Note that since for black holes $M>a$, this implies that $b$ is imaginary, which is not an issue for the MSM spacetime \cite{MSM}.} If the quadrupole deviation parameter is defined as
\begin{align}\label{eq:QuadDeform}
 \delta{\cal Q}:={\cal Q}-{\cal Q}_\textrm{Kerr}=\frac{M\left(M^2+b^2-a^2\right)^2}{4(a^2-2 a\,b+b^2-M^2)}\, ,
\end{align}
then for $\delta{\cal Q}>0$ the MSM describes a more prolate bumpy black hole than the Kerr one and for $\delta{\cal Q}<0$ a more oblate one. For $b=0$ the Tomimatsu Sato $\delta=2$ solution is retrieved \cite{MSM} and $\displaystyle \delta{\cal Q}=-\frac{M}{4}(M^2-a^2)$.
 
For presenting the spacetimes deviating from Kerr, it is useful to introduce the \ix{Weyl set of coordinates}, which relates to the Boyer-Lindquist coordinates as follows:
\begin{align}\label{eq:WP2Str}
    \rho=\sqrt{\Delta} \sin{\vartheta},\,z=(r-M) \cos{\vartheta}
\end{align}
Note that in the Weyl set of coordinates the event horizon $r_H=M+\sqrt{M^2-a^2}$ for Kerr lays at $\rho=0$ and stretches along $z$ in the interval $\left[-\sqrt{M^2-a^2},\sqrt{M^2-a^2}\right]$, i.e. the horizon is reduced to a line segment along the $z$-axis. Obviously, this set of coordinates does not cover the spacetime inside the event horizon.\footnote{Even though it is possible to use imaginary values of the coordinates to reach the black hole interior, see Ref. \cite{basovnik2016geometry}.} The line element in Weyl coordinates reads
\begin{align}\label{eq:WLnEl}
  ds^2  =  -e^{2 \nu}(dt-\gamma d\varphi)^2 %\nonumber \\
        +  e^{-2 \nu} \left[ e^{2\psi} (d\rho^2+dz^2)+\rho^2 d\varphi^2 \right]\,.
 \end{align}

For the reduced MSM spacetime, which is used for the numerical examples in the article, the metric functions read:  
 \begin{align} \label{eq:MetrFunc}
  e^{2 \nu} &= \frac{{\cal E}}{({\cal E}+R\, P+(v^2-1) S\, T)} \, , \\
  e^{2\psi} &=\frac{ {\cal E}}{16 \kappa^8 (u^2-v^2)^4} \, ,\nonumber \\
  \gamma &= \frac{ (v^2-1)(R\, T-\kappa^2 (u^2-1) S\, P)}{{\cal E}} \, , \nonumber \\
  {\cal E} &= R^2+\kappa^2 (u^2-1) (v^2-1) S^2\, , \:
  S  =  -4 {(a-b)[\kappa^2(u^2-v^2)+2 \delta v^2]+v^2 M^2 b}\, , \nonumber \\
  P & =  2 \{\kappa M u [(2 \kappa u+M)^2-2 v^2 (2 \delta +a b -b^2)  -  a^2+b^2 ]-2 v^2 (4 \delta d-M^2 b^2)\}\quad, 
   \nonumber \\
  R & =  4 [\kappa^2 (u^2-1)+\delta (1-v^2)]^2 
  +  (a-b) [(a-b)(d-\delta)-M^2 b](1-v^2)^2\quad,\nonumber \\
  T & =  4(2 \kappa M b u+2 M^2 b)[\kappa^2 (u^2-1)+\delta (1-v^2)]
     +  (1-v^2)\{(a-b)(M^2 b^2 -4 \delta d)  \nonumber \\
    & -  (4 \kappa M u+2 M^2) [(a-b)(d-\delta)-M^2 b]\}\, ,
 \end{align}
  where
 \begin{align} \label{eq:kdd}
  \kappa =\sqrt{d+\delta}\, ,\: \delta = -\frac{M^2 b^2}{M^2-(a-b)^2} \, , \: d = \frac{1}{4}[M^2-(a-b)^2]\,. 
 \end{align}
 Note that all the metric functions are expressed in generalized spheroidal coordinates $u,~v$, while the line element~\eqref{eq:WLnEl} is written in the Weyl coordinates $\rho,~z$. The transformation between them reads
 \begin{align} \label{eq:TrCylPS}
  \rho =\kappa \sqrt{(u^2-1)(1-v^2)} \, ,\: z=\kappa u v\, .
 \end{align}
 For $b^2=a^2-M^2$, it can be shown that $\kappa=\sqrt{M^2-a^2}$, $v=\cos\vartheta$ and $u=(r-M)/\kappa$.  

\begin{figure}[ht] 
    \centering
    \includegraphics[width=0.49\textwidth]{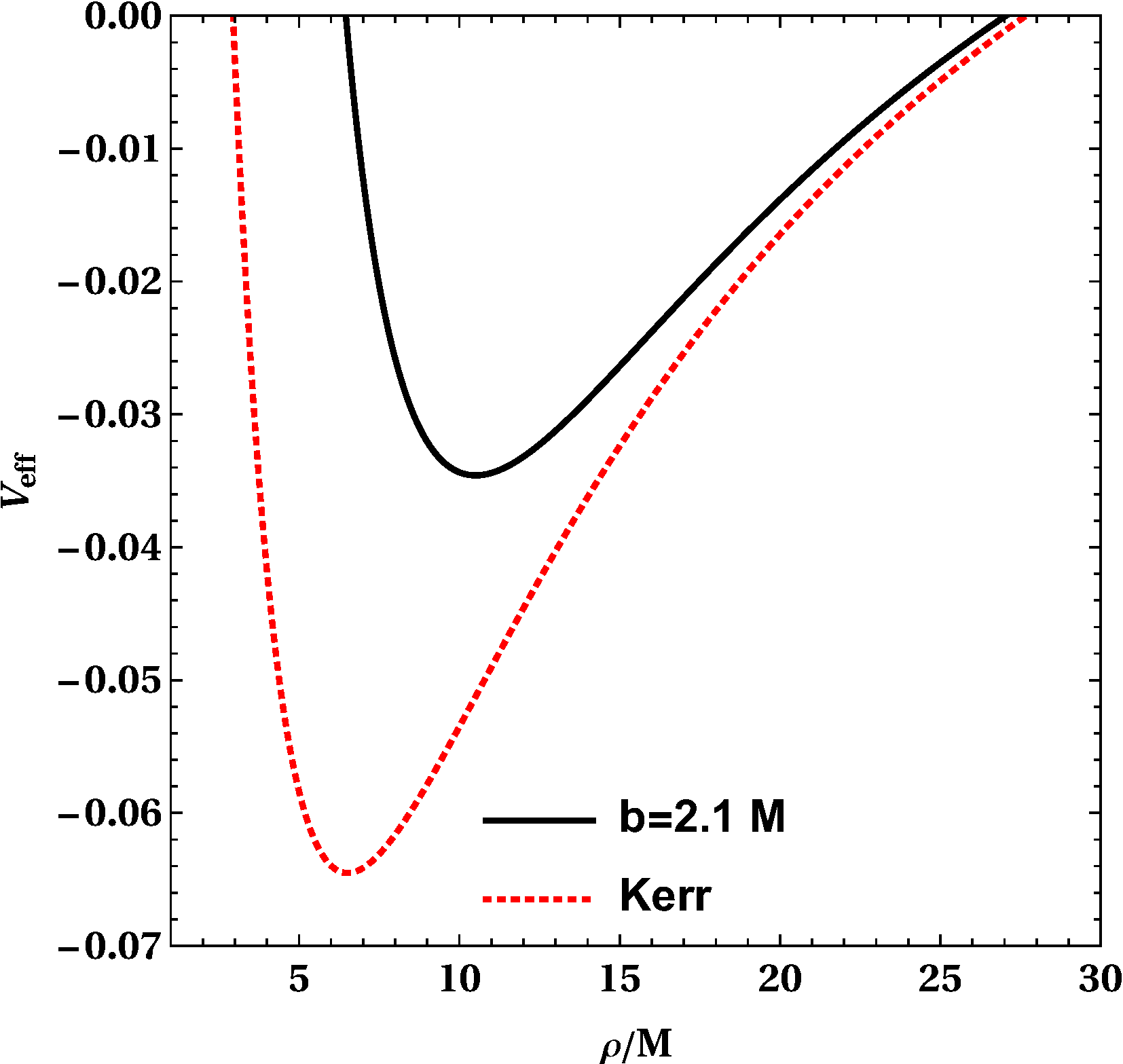}   
        \includegraphics[width=0.49\textwidth]{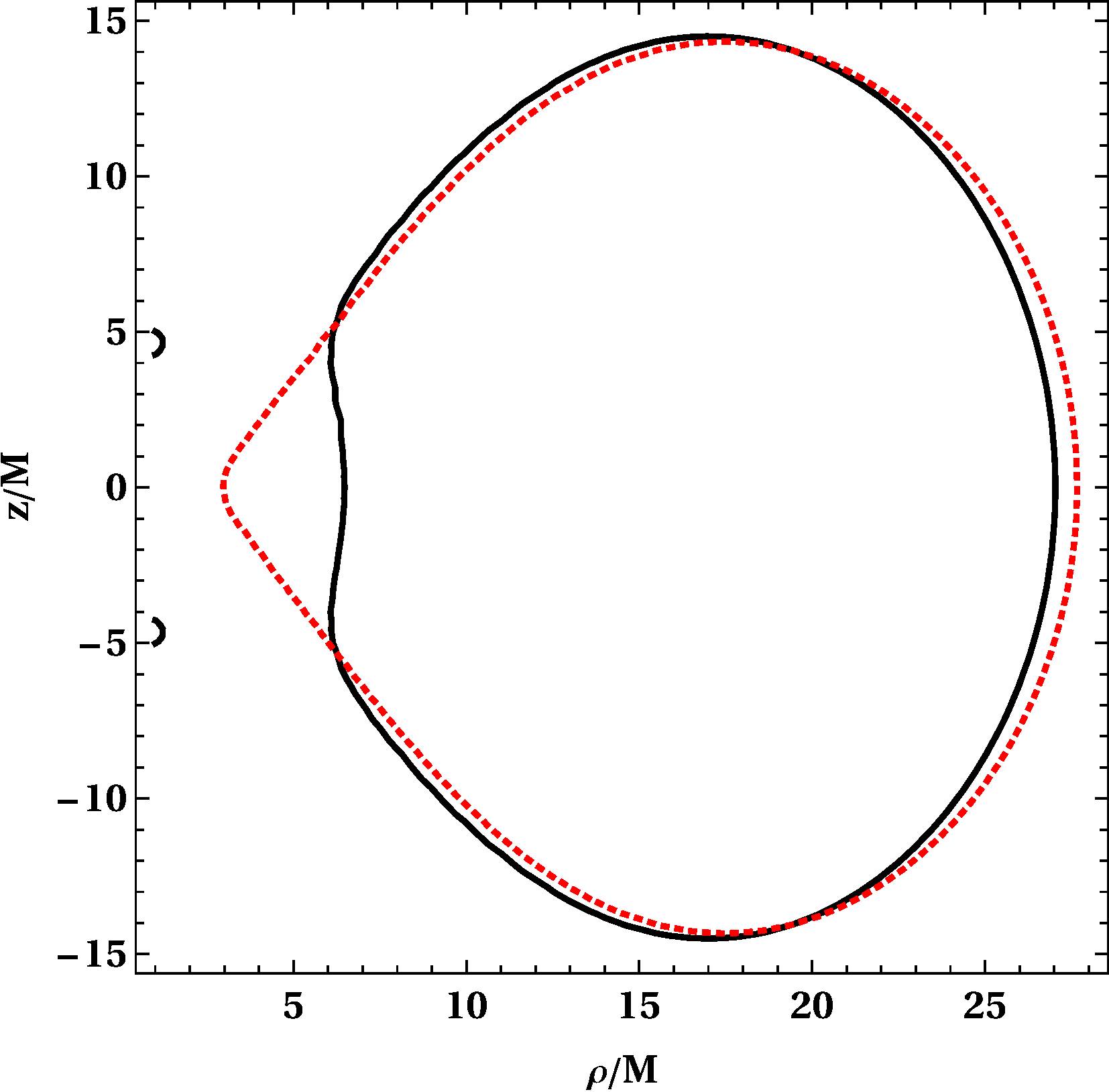}  
    \caption{Left panel: The effective potential when $z=0$ for the Kerr spacetime with $a=0.999 M$ and the specific orbital parameters set to $E=0.97$ and $Lz=3 M$ (red dotted curve) and the respective effective potential for MSM spacetime when $b=2.1 M$ (black curve). Right panel: The curve of zero velocity $V_\textrm{eff}=0$ for the cases shown in the left panel.}
    \label{fig:Veff}
\end{figure}

A useful concept to study the geodesic motion in a spacetime~\eqref{eq:WLnEl} is that of the effective potential. If we take the Hamiltonian function~\eqref{eq:HamGeo} and divide it with the square of the test particle, we arrive to 
\begin{align} \label{eq:effectPot}
   -\left( g_{\rho\rho}\dot{\rho}^2 +g_{zz}\dot{z}^2\right)= V_\textrm{eff}=(1+g^{tt}E^2-2g^{t\varphi}E~L_z+g^{\varphi\varphi} L_z^2)\, ,
\end{align}
where $\displaystyle \dot{x}^\mu=\frac{d x^\mu}{d \tau}$ is the four-velocity, while $E=-p_t/\mu$ and $L_z=p_\varphi/\mu$ are the specific energy and angular momentum respectively, i.e. the energy and angular momentum per unit mass $\mu$.\footnote{For simplicity, we keep the same symbols $E,L_z$ as for the respective quantities not divided by particle rest mass. Which type is used each time is either stated explicitly or implied by the units.} The lhs of Eq.~\eqref{eq:effectPot} is a strictly non-positive quantity, since $g_{\rho\rho}=g_{zz}>0$ outside the horizon, implying that geodesic motion is allowed only if $V_\textrm{eff}\leq 0$ (Fig.~\ref{fig:Veff}). The $V_\textrm{eff}=0$ limit defines the curve of zero velocity, i.e the limit along which a geodesic has zero radial and polar velocity. On the equatorial plane $z=0$ bounded orbits are defined by two radii $\rho_1<\rho_2$, for which $V_\textrm{eff}(\rho_1)=V_\textrm{eff}(\rho_2)=0$ (Fig.~\ref{fig:Veff}). For a circular orbit it holds that $V_\textrm{eff}=\frac{\partial V_\textrm{eff}}{\partial \rho}=0$, while the stability of circular orbits is defined by the sign of $\frac{\partial^2 V_\textrm{eff}}{\partial \rho^2}$: positive for the stable ones and negative for the negative. Note that the definition of the effective potential is not unique. For example, alternatively one can define the following expression as effective potential
\begin{align}\label{eq:EffPotA}
    V_{\rm eff} (\rho,z;L_z) = E(\dot{\rho}=\dot{z}=0,L_z) = \frac{g^{t\varphi}}{g^{tt}} L_z +\sqrt{ \left[\left(\frac{g^{t\varphi}}{g^{tt}}\right)^2 - \frac{g^{\varphi \varphi}}{g^{tt}}\right]L_z^2- \frac{1}{g^{tt}}}
\end{align}
as was done for Fig.~\ref{fig:Schwarzport}. Despite the different definitions, the dynamics defined by any of the effective potentials are obviously the same.

\subsubsection{External matter deformation}\label{sec:ext_matt}

Another way to introduce deformations to the black hole field to cause chaos is by surrounding it by additional matter \cite{Semerak15}. The gravitating matter can correspond to accretion disks or rings, clouds of dust, or matter halos. Exact solutions for spinning black holes surrounded by matter are rare \cite{Bicak93,Neugebauer93}, however, solutions valid to linear order in the matter perturbation can be constructed using the Teukolsky equation and the so-called Chrzanowski \& Kegeles formalism \citep[see the example and references in][]{sano2014gravitational}. On the other hand, in the case of a Schwarzschild black hole many idealized exact solutions can be found \cite{Lemos94,Semerak03,Semerak20}. The effect of the self-gravitating matter on the geodesic motion around a Schwarzschild black hole is that the total angular momentum ceases to be a constant of motion, since the spherical symmetry is lost and reduced usually to axisymmetry. The loss of the fourth integral of motion, give rise to non-integrability and chaos \cite{Semerak15}.

Let us take a look at how such a superposition of a Schwarzschild black hole with additional matter source looks like. For a static and axially symmetric spacetime $\gamma=0$ and $\nu,~\psi$ are functions only of $\rho$ and $z$ in the line element~\eqref{eq:WLnEl}. The vacuum Einstein's field equations then reduce to a Laplace equation
\begin{align}\label{eq:Laplace}
  \rho \left( \frac{\partial^2 \nu}{\partial \rho^2}+\frac{\partial^2 \nu}{\partial z^2}\right)+\frac{\partial \nu}{\partial \rho}=0
\end{align}
and a line-integral equation
\begin{align}\label{eq:linInt}
   & \frac{\partial \psi}{\partial \rho} = -\rho\left(\left(\frac{\partial \nu}{\partial z}\right)^2-\left(\frac{\partial \nu}{\partial \rho}\right)^2\right)\,,\\
   & \frac{\partial \psi}{\partial z} = 2\rho\frac{\partial \nu}{\partial z}\frac{\partial \nu}{\partial \rho}\,.
\end{align}

The Laplace equation~\eqref{eq:Laplace} implies that $\nu$ "potentials" can be added linearly like in Newtonian theory, which, however, does not hold for the function $\psi$. Namely, if we have source described by $\nu_1,~\psi_1$ and another source described by $\nu_2,~\psi_2$, then the superposition of these sources is given by the potential $\nu=\nu_1+\nu_2$ and the function $\psi=\psi_1+\psi_2+\psi_\textrm{int}$. The interaction term $\psi_\textrm{int}$ can be obtained by integrating
\begin{align}\label{eq:intfunc}
&\frac{\partial \psi_\textrm{int}}{\partial \rho} = -2\rho\left(\frac{\partial \nu_1}{\partial z}\frac{\partial \nu_2}{\partial z}-\frac{\partial \nu_1}{\partial \rho}\frac{\partial \nu_2}{\partial \rho}\right) \,,\nonumber \\
&\frac{\partial \psi_\textrm{int}}{\partial z} = 2\rho\left(\frac{\partial \nu_1}{\partial \rho}\frac{\partial \nu_2}{\partial z}+\frac{\partial \nu_1}{\partial z}\frac{\partial \nu_2}{\partial \rho}\right)\, .
\end{align}
In particular, a Schwarzschild black hole is described by the functions
\begin{align}
    \nu_\textrm{S} &=\frac{1}{2}\ln{\frac{d_{+}+d_{-}-2\,M}{d_{+}+d_{-}+2\,M}} =
    \frac{1}{2}\ln{\left(1-\frac{2M}{r}\right)} \, ,\\
    \psi_\textrm{S} &=\frac{1}{2}\ln{\frac{(d_{+}+d_{-})^2-4\,M^2}{4\,d_{+}d_{-}}} =
    \frac{1}{2}\ln{\frac{r(r-2M)}{(r-M)^2-M^2\cos^2(\vartheta)}}\, ,
\end{align}
where 
\begin{align}
    d_{\pm}=\sqrt{\rho^2+(z\pm M)^2}=r-M\pm M \cos{\vartheta}\, .
\end{align}
For the respective functions describing the surrounding matter the interested reader is referred to the literature, see, e.g., \cite{Lemos94,Semerak03,Semerak15,Semerak20}. 

Let us now discuss a simple case of a black hole immersed in the field of an axisymmetric faraway halo or ring of matter of characteristic distance $R$ from the black hole, and mass $\mathcal{M}$. In the interior of this halo or ring, the leading-order effect will be the quadrupolar tidal. The metric functions for this tidal field read \cite{doroshkevich1966}
\begin{align}
    \nu_Q &=-\frac{1}{4}Q\left(\rho^2-2\,z^2\right)\, ,\nonumber\\
    \psi_Q &= \frac{\rho^2}{2}\left(\frac{\rho^2}{8}-z^2\right)Q^2 \,,
\end{align}
where $Q \sim \mathcal{M}/R^3$ is the quadrupolar parameter and it is easy to check that the metric functions satisfy Eqs.~\eqref{eq:Laplace},~\eqref{eq:linInt}. Obviously, this space-time is physically valid only at distances much smaller than $R$ from the center and, as a consequence, it is not even asymptotically flat. However, for the purposes of our didactic example this metric is sufficient. 

The interaction function $\psi_\textrm{int}$ for the above sources is derived from integrating Eq.~\eqref{eq:intfunc} and reads
\begin{align}
    \psi_\textrm{int}=-\frac{1}{2}\left((z+M)d_-+(M-z)d_+)\right)Q\,.
\end{align}
Even if the functions of the superposition $\nu_\textrm{SQ}=\nu_\textrm{S}+\nu_\textrm{Q}$ superpose linearly, the metric functions are an infinite series in $Q$. However, we are able to write the Hamiltonian of the geodesic motion in this space-time as
\begin{align}
    H_{\rm SQ} =& H_{\rm S} + Q h_{Q} + \mathcal{O}(Q^2) \,,\\
    \begin{split}
   h_{Q} =&  - \frac{1}{4} \left(\rho^2 - 2 \, z^2\right) \left[e^{-2 \nu_\textrm{S}}p_t^2 + e^{2 \nu_\textrm{S}} \left( e^{-2 \psi_{\rm S}}(p_\rho^2 + p_z^2) + \frac{p_\varphi^2}{\rho^2} \right) \right] \\ 
         & -\frac{1}{2}\left((z+M)d_-+(M-z)d_+)\right) e^{2 \nu_{\rm S} - 2 \psi_{\rm S}} (p_\rho^2 + p_z^2)\,.
\end{split}
\end{align}
In other words, we can treat the external tidal field as a perturbation to the Hamiltonian, which is subject to the theory discussed in the previous sections of this paper. One can then use analytical perturbation methods such as the so-called Melnikov integral to see whether this perturbation will cause non-integrability along the homoclinic orbits in Schwarzschild space-time \citep[see][]{polcar2019free}.

\subsection{Spinning particle}

Perturbing the background spacetime is not the only way the motion can become non-integrable. Another way non-integrability can arise is when we take the internal multipole structure of the test body into account \cite{Suzuki97}. The \ix{Mathisson-Papetrou-Dixon (MPD) equations} \cite{Mathisson37,Papapetrou51,Dixon74} describe the motion of an extended test body on a curved spacetime. If the multipole expansion of the body is truncated to the pole-dipole approximation, then the body is effectively to a \ix{spinning particle} and the MPD equations read
\begin{align}
    \frac{D~P^\mu}{d \lambda} &= -\frac{1}{2}{R^\mu}_{\nu\rho\sigma}U^\nu S^{\rho\sigma} \:,\\
    \frac{D~S^{\mu\nu}}{d \lambda} &=  P^\mu U^\nu - P^\nu U^\mu \: \label{eq:mpdspin},
\end{align}
where $S^{\mu\nu}$ is the spin-tensor, $P^\mu$ is the four-momentum, $U^\mu = \d x^\mu/\d\lambda $ is a tangent vector and the $\displaystyle\frac{D}{d \lambda}$ denotes a covariant derivative with respect to an affine parameter $\lambda$. The MPD equations are underdetermined; one has to add four constraints in order to evolve the system. One constraint comes from defining the affine parameter, while the other three come from a spin supplementary condition (SSC)
\begin{align} \label{eq:SSC}
    V_\mu S^{\mu\nu}=0\, 
\end{align}
fixing the centre mass of the body, where $V^\mu$ is a future-oriented time-like vector. If the affine parameter is the proper time $(U^\mu U_\mu=-1)$, then the MPD equations can be recovered from a Hamiltonian for certain SSCs and its constant value expresses the conservation of the particle mass, similar to the case of time-like geodesics \cite{Witzany19}. The spin of the particle introduces one additional degree of freedom as compared to the geodesic motion, even though there can be additional ``gauge'' degrees of freedom involved in the evolution in some cases \cite{Witzany19}. The strength of the deviation of the spinning particle motion from geodesic motion as well as the coupling of the spin degree of freedom to the orbit are governed by the spin magnitude $S$ defined by
\begin{align}
    S^2=\frac{1}{2}S^{\mu\nu}S_{\mu\nu}\,.
\end{align}

For each continuous symmetry of the spacetime background with a corresponding Killing vector $\zeta^\mu$, this system admits an integral of motion 
\begin{equation}\label{eq:int_mpd}
    C(\zeta) = P_{\sigma}\zeta^{\sigma} - \frac{1}{2}\zeta_{\rho;\sigma}S^{\rho\sigma} \:.
\end{equation}
For the Schwarzschild background, there are 4 Killing vectors corresponding to rotations and time translations
\begin{align}
    \zeta_{(t)} &= \pder{}{t} \:, \label{eq:StaSym} \\
    \zeta_{(z)} &= \pder{}{\varphi} \: , \label{eq:AxiSym} \\
    \zeta_{(x)} &= -\sin\varphi\pder{}{\vartheta} - \cos\varphi\cot\vartheta\pder{}{\varphi} \:, \\
    \zeta_{(y)} &= \cos\varphi\pder{}{\vartheta} - \sin\varphi\cot\vartheta\pder{}{\varphi} \:  .
\end{align}
These generate the following integrals of motion: the energy $E:=-C(\zeta_{(t)})$ and the three components of the total angular momentum $J_x:=C(\zeta_{(x)}),~J_y:=C(\zeta_{(y)}),~J_y:=C(\zeta_{(z)})$. From the three components of the angular momentum we can obtain only at most two integrals in involution, since they have they Poisson-commute with the commutation relations of the generators of rotations, e.g., $\{J_x,J_y\}=J_z$. We can choose the two integrals in involution for instance as $J^2,J_z$. In the Kerr case, however, there are only the Killing vectors $\zeta_{(t)},~\zeta_{(z)}$ and the respective integrals of motion $E,J_z$ and no exact generalization of $J^2$.

From now on, we are going to constrain the discussion on the case when $V^\mu=p^\mu$, which is the Tulczyjew-Dixon (TD) SSC. In the case of the Schwarzschild background the extra degree of freedom comes without a respective integral of motion. Using $E,J^2,J_z$ the motion of the spinning particle on the Schwarzschild background can be reduced to a two-degree system \cite{Zelenka20}. In the case of a Kerr background, the Carter constant is no longer constant for a spinning particle. Hence, by using the $E$ and $J_z$ the motion of the spinning particle on a Kerr background can be reduced to a three  degree of freedom system \cite{Witzany19}. In both the Kerr and Schwarzschild cases, the spin of the test particle leads to a weakly non-integrable system, however, in the Kerr case one has to deal with more degrees of freedom leading to more complex dynamics \cite{LG16}. 

From the MPD equations the spin is naturally identified as the perturbation parameter. One might expect that once the perturbation is present the system should become non-integrable. However, the picture can be more complicated. For example, in the linear-in-spin approximation of MPD equations on a Kerr background at least for the TD SSC integrability can be approximately recovered up to $\mathcal{O}(S^2)$ \cite{Witzany19b}, since two other integrals are conserved to linear order in spin \citep{ruediger1,ruediger2}. Namely, these two integrals are a Carter-like quantity and a quantity corresponding to the projection of the orbital momentum on the spin \cite{Witzany19b}. Hence, when an integrable system is perturbed, the system might hold some of its integrable features up to an order in its expansion with respect to the perturbation parameter.

Let us see what is the expected contribution of a spin induced prolonged resonance crossing to an inspiral. The width of the spin induced prolonged resonances grows linearly with the value of the dimensionless spin $\sigma=\frac{S}{\mu M}$ \cite{Zelenka20}. Taking into account the relation~\eqref{eq:pert_to_wid}, this implies that the crossing will last $\sqrt{\sigma^2}/q$ cycles (see section~\ref{sec:addpet}). Since the dimensionless spin $\sigma$ of a secondary compact object, like a black hole or a neutron star, is of the same order as to the mass ratio $q$. Hence, the spin induced resonance will dominate at least over $\mathcal{O}(1)$ cycles during an inspiral.

\section{Impact of non-integrability on extreme-mass-ratio systems}

We would now like to demonstrate in a qualitative model what kind of effect can the non-integrability of geodesics have on an inspiral in the space-time. Recall that in the simplest approximation an EMRI can be considered as an isolated binary system, in which the secondary body is regarded as a non-extended test particle, while the primary body is a Kerr black hole defining the background in which the secondary moves. In this picture the secondary drifts adiabatically from a geodesic to a geodesic trajectory due to gravitational radiation reaction, essentially tracing out the phase space of integrable Kerr geodesics. In the long run the motion depends on the slowly dissipating action variables, which in physical terms translate to losses in energy, orbital angular momentum $L_z$ and Carter constant $\mathcal{K}$. Seen from the lens of dynamical-systems theory, the EMRI in the above setup can be viewed as a regular dissipative dynamical system where the phase space is separable into the space of actions and the dependent phases, which do not feed back into the evolution in the action space at all.

In a more accurate EMRI approximation, the secondary body is extended and the self-force has also a conservative part affecting the immediate motion. The full description of self-force to the first order terms has been reached only recently for the Kerr black hole background \cite{van2018gravitational,Barack19} and it is very difficult to achieve for other spacetimes like the bumpy black holes. Moreover, the issue is that the self-force computation in a given space-time represents a huge investment, and it is not clear from astrophysical observations for which metric in particular should one carry this computation out. 
Therefore, to simulate the dissipative part one can employ simple qualitative formulas such as the quadrupole formula or some appropriately modified kludge prescriptions to gain qualitative insight into the effect.

As long as an inspiral crosses regular parts of the phase space, the adiabatic approximation holds and, from a dynamical point of view, one cannot tell whether the system is globally integrable or not. Discrepancies will arise only when the inspiral reaches a resonance. Each strong resonance should introduce to the inspiral a phase shift of the order of $\sqrt{\epsilon}/q$, where $q$ is the mass ratio and $\epsilon$ the perturbation parameter. Therefore, before including the dissipation into a perturbed system deviating from the simplest approximation, it is important to estimate how the perturbation parameter $\epsilon$ leads to non-integrability, i.e. how resonances and, hence, chaos grow with respect to $\epsilon$. Analytical estimates of the resonant growth are quite difficult, since they require handling the geodesic motion in action-angle coordinates. Thus, we will now present how one can investigate this question numerically.

 \subsection{\ix{Resonance growth}}

\begin{figure}[ht] 
    \centering
    \includegraphics[width=0.49\textwidth]{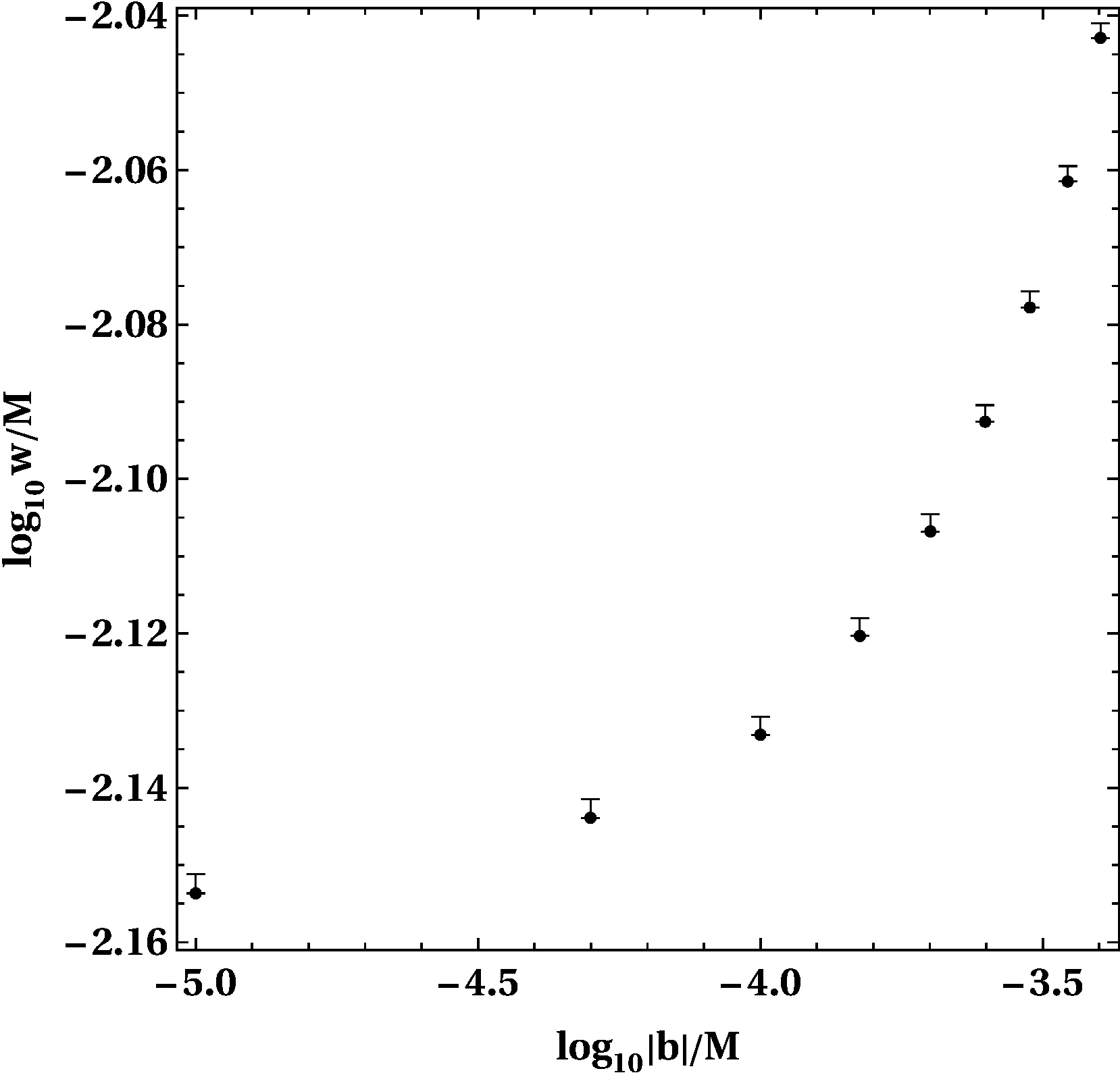}  
    \includegraphics[width=0.49\textwidth]{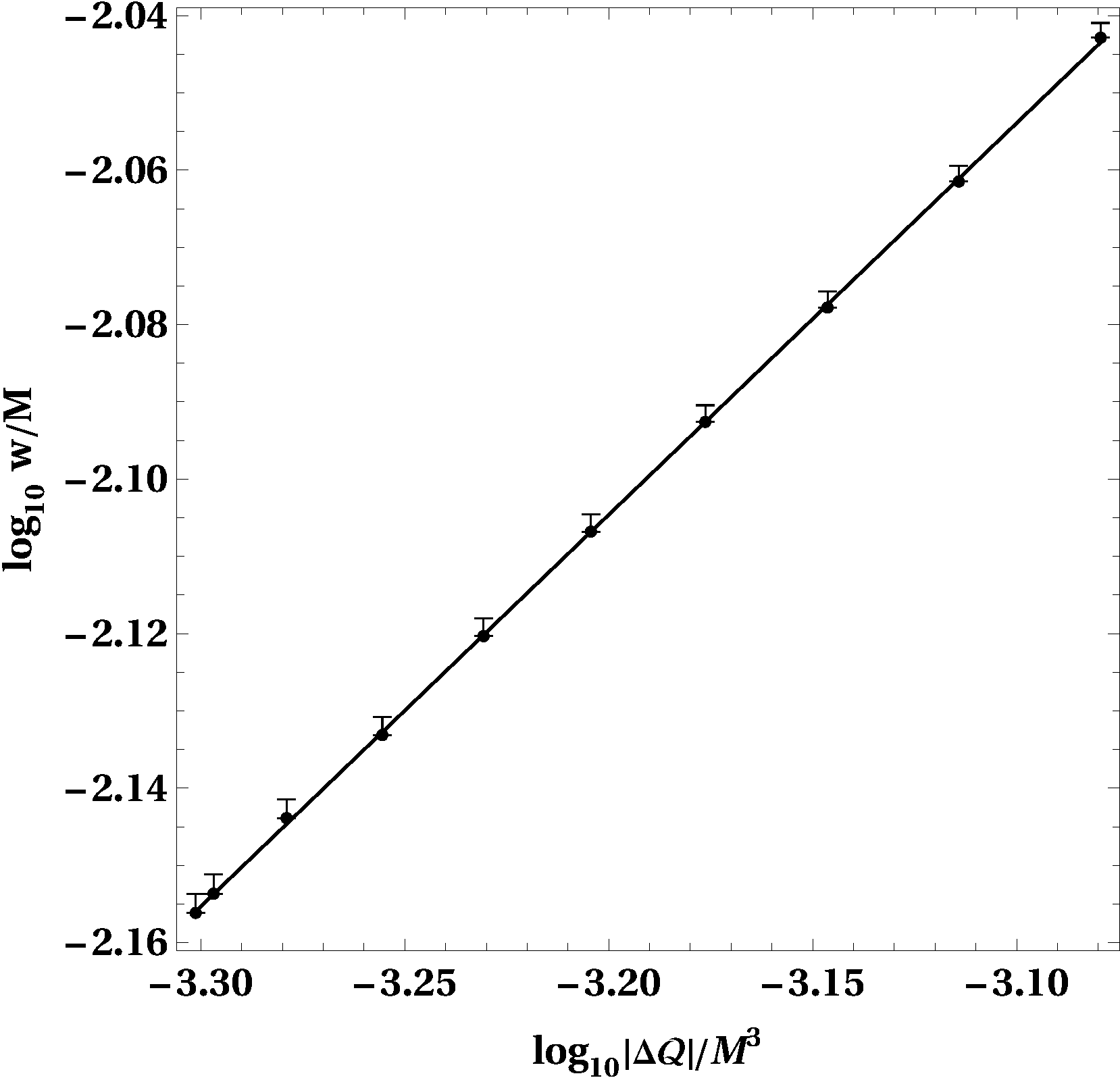}    
    \caption{Both panels show how the resonance $1/3$ grows, the left panel with respect to the parameter $b$ and the right panel with respect to the quadrupole deviation from Kerr $\Delta\mathcal{Q}$. Spin is set to $a=0.999 M$ and the orbital parameters to $E=0.97$ and $L_z=2 M$. On Poincar\'{e} sections similar to the right panel of Fig.~\ref{fig:sec_rot} rotation curves have been created along the $\dot{\rho}=0$ with a step $\Delta \rho=2~10^{-5} M$. By measuring the plateaus on the rotation curve the width of the resonances have been found with accuracy $2~\Delta\rho$. By applying a linear fit on the points shown in the right panel, the inclination has been found to be $0.5069\pm 0.018$, which indicates that the width of the resonance is proportional to $\sqrt{\Delta \mathcal{Q}}$.}
    \label{fig:reswidth}
\end{figure}

\begin{table}[]
    \centering
    \begin{tabular}{c||c|c|c|c|c|c|c|c|c|c}
      b $(10^{4}/M)$  & $-4$ & $-3.5$ & $-3$ & $-2.5$ & $-2$ & $-1.5$ & $-1$ & $-0.5$ & $-0.1$ & $0$ \\ \hline
     width $(10^3/M)$  & $9.06$ & $8.68$ & $8.36$ & $8.08$ & $7.82$ & $7.58$ & $7.36$ & $7.18$ & $7.02$ & $6.98$  
    \end{tabular}
    \caption{The values of $b$ used to produce Fig.~\ref{fig:reswidth} and the obtained respective widths of the $1/3$ resonance. }
    \label{tab:width}
\end{table}

Assume that we have enough integrals to reduce a non-integrable system to two degrees of freedom. In such a system we can use Poincar\'{e} sections and rotation curves to detect a resonance and to follow its width modification as the perturbation parameter $\epsilon$ changes. This can be achieved either by looking for the unstable points of the resonance in order to measure the angle between the asymptotic manifold branches as done in \cite{Zelenka20} or by looking for stable points in order to measure the width of the plateaus on the rotation curves as done to obtain Fig.~\ref{fig:reswidth}. In particular, for each of the $b$ parameter values shown in Table~\ref{tab:width} the width $w$ of the plateau was measured. Since the perturbation parameter $\epsilon$ is expected to be in a power law relation with the parameter of the system deviating it from integrability, plotting the width $w$ with respect to this parameter should give a straight line in a logarithmic plot. The left panel of Fig.~\ref{fig:reswidth} shows that $b$ is not that parameter, since the points on the plot do not correspond to a line. On the other hand, the right panel indicates that the quadrupole deviation parameter $\Delta\mathcal{Q}=\epsilon$, since the points fit well a line with an inclination $\approx 0.5$ and there is Eq.~\eqref{eq:pert_to_wid} relating the width of the resonance with the perturbation parameter. 

\subsection{\ix{Prolonged resonances}}

\begin{figure}[ht] 
    \centering
    \includegraphics[width=0.49\textwidth]{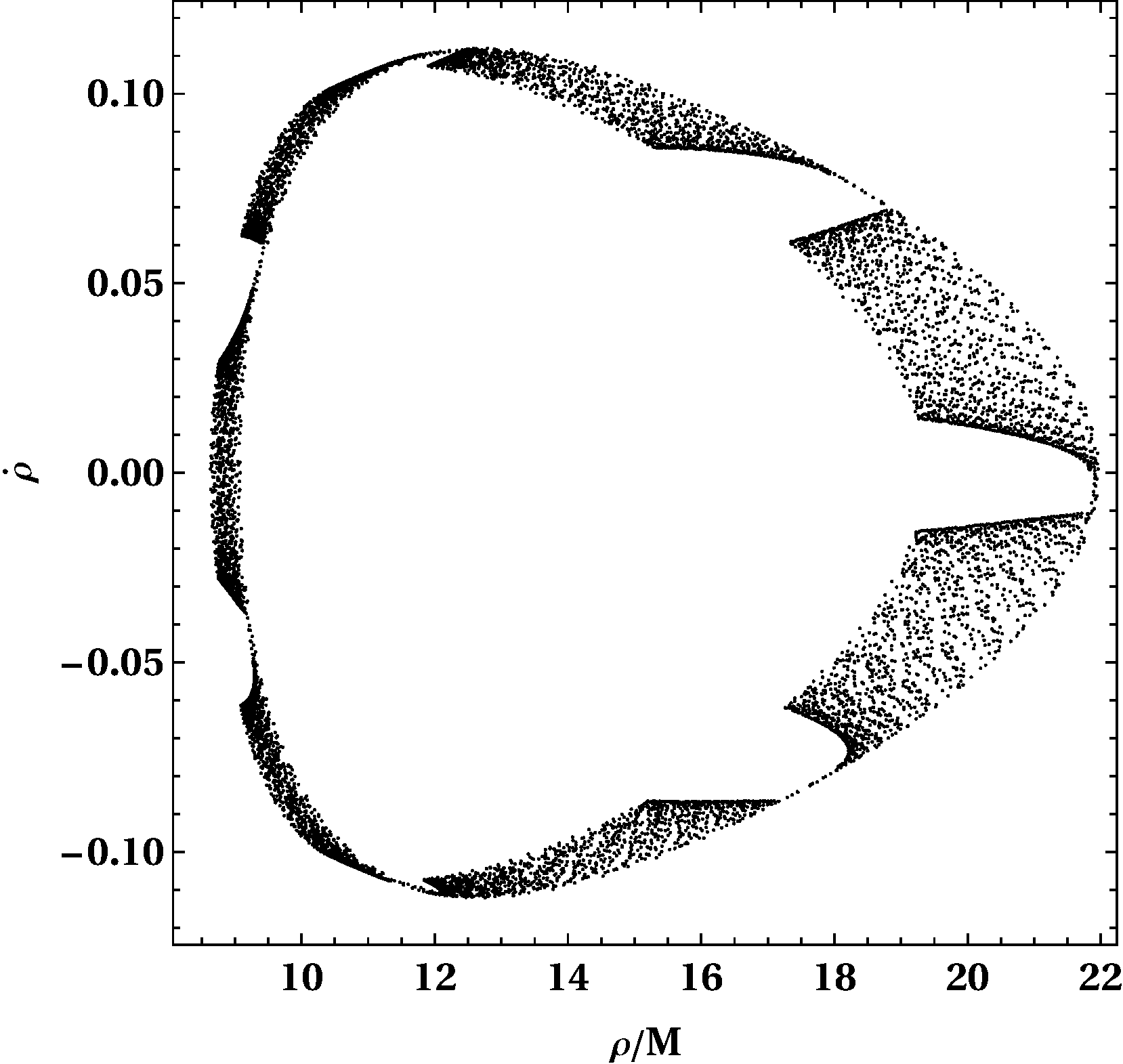}  
    \includegraphics[width=0.49\textwidth]{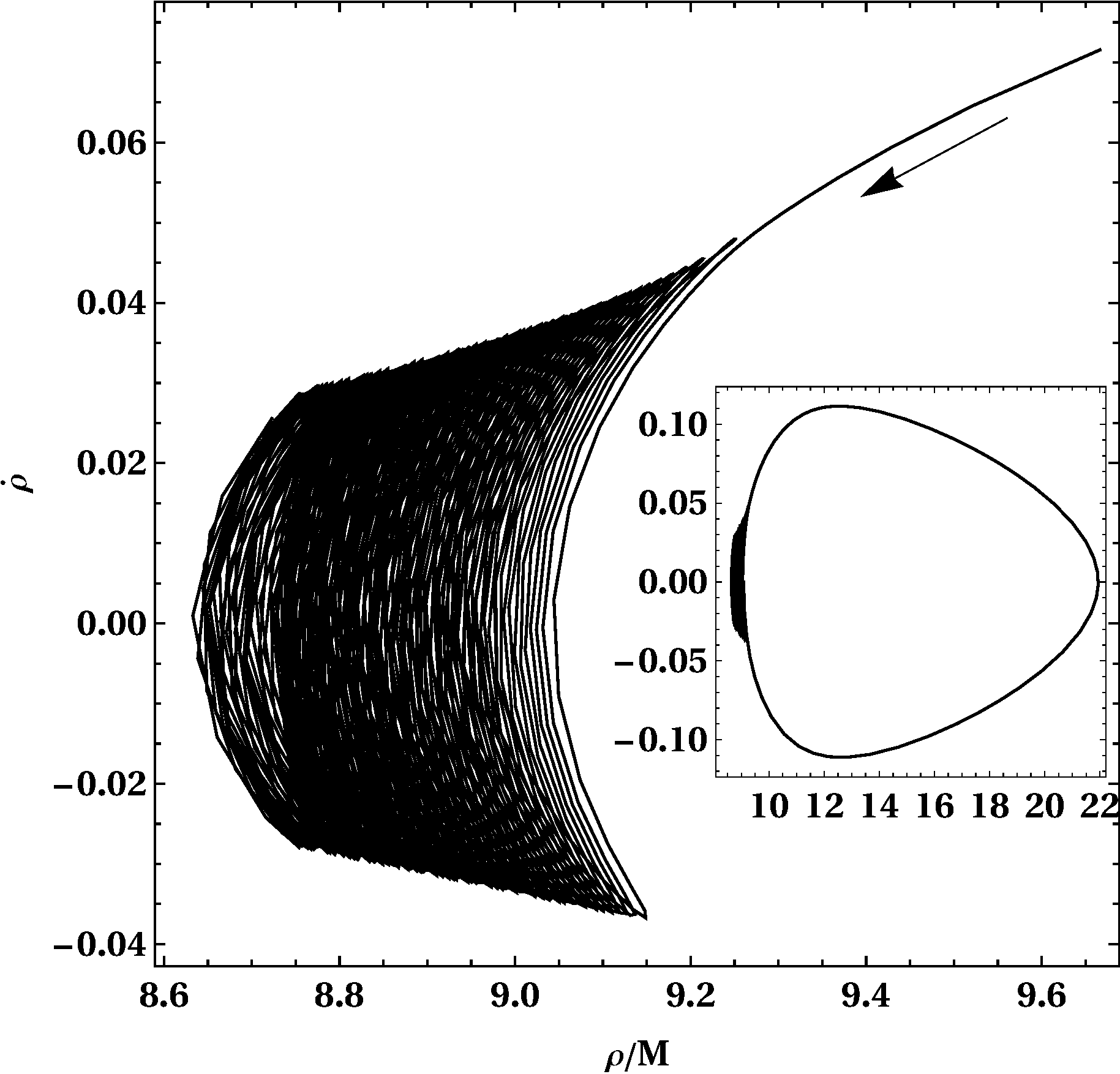}    
    \caption{An inspiral with mass ratio $q=10^{-4}$ going through $6/7$ resonance on a MSM spacetime with $M=1,~a=0.999~M,~b=2.1~M$. The inspiral starts from the equatorial plane with $\rho=9.0790461 M,~\dot{\rho}=0$ with energy $E|_g=0.97 \mu$ and angular momentun $L_z|_g=3 \mu M$. This initial condition is not unique, similar inspiraling behavior can be found in the interval $9.07898~M\lessapprox \rho \lessapprox 9.07902~M$ on the equatorial plane.  Left panel: The inspiral depicted on a section lying on the equatorial plane. Right panel: Detail from the stroboscopic depiction of left panel panel, the full picture is shown as an inset.}
    \label{fig:res67}
\end{figure}

Now that we have established the relation between the quadrupole deviation parameter and the perturbation parameter, let's see what happens to the inspiral during a resonance crossing. For that we follow the recipe given in \cite{LGAC10}. Namely, we have used modified kludge formulae as in \cite{LGAC10} to calculate the energy $d E/dt|_g$ and angular momentum $dL_z/dt|_g$ fluxes of the secondary on a geodesic trajectory. Then we subtract these fluxes from the energy $E|_g$ and angular momentum $L_z|_g$ in a linear approximation, i.e. $E(t)=E|_g-\frac{d E}{dt}|_g t$ and $L_z(t)=L_{z}|_g-\frac{d L_z}{dt}|_g t$, to introduce dissipation into the system. Note we do not attempt to introduce dissipation to the other components of angular momentum or some generalization of the Carter constant in the MSM space-time. We have employed the above described procedure on the $6/7$ resonance shown in the right panel of Fig.~\ref{fig:sec_rot} and on the $1/3$ resonance for $b=10^{-4}M$ (one of the cases of Fig.~\ref{fig:reswidth}). 

Starting on a given surface of section with a resonance, it is not easy to guess where to place an initial condition on the section so that it crosses the resonance. This is because the section is actually constructed at constant $E,L_z$ and the dissipation thus makes us drift between various sections. This can also be stated in orbital parameter terms, it is not clear whether during the evolution of the inspiral a resonance lying at higher inclinations and eccentricities (at fixed $E,L_z$) in the phase space than the inspiraling body will catch up with the inspiral or whether the inspiraling body starting from higher inclinations and eccentricities (at fixed $E,L_z$) will catch up with the resonance and cross it. In our numerical investigations and in \cite{LGAC10} the resonance crossings takes place only when the initial conditions are set between the resonance and the main island of stability.   

Another unclear aspect of prolonged resonance crossings is the time that will the inspiral spend in the resonance. For example, the $1/3$ resonance crossing examined here and the $2/3$ resonance crossing discussed in \cite{LGAC10} indicate that the inspiral will enter and leave the resonance in finite time. The amount of time spent in the resonance varies from initial condition to initial condition and the only way to tell why would be to carry out an analysis as sketched in Section \ref{sec:resopass}. A more puzzling case is that there appears to be cases that the inspiral gets trapped in the resonance. In particular, when we examined the crossings of the $6/7$ resonance, we have found that there are initial conditions producing  cases that the inspiral enters the resonance, but does not seem to be able to leave it (see Fig.~\ref{fig:res67}).

Let us examine more carefully the trapping case. The left panel of Fig.~\ref{fig:res67} shows the intersections of the inspiraling orbit and the equatorial plane on the $\rho,~\dot{\rho}$ surface, where $\dot{\rho}=p^\rho/\mu$ is the $\rho$ component of the four-velocity. Note that such a plot would be a Poincar\'{e} section if not for the dissipation. Having this in mind we borrow the terminology for the description of the plot from the conservative counterpart. The plot shows an inspiral with mass ratio $q=10^{-4}$ starting from a KAM torus (a continuous contour) before it enters the resonant island of stability (7 distinct regions in the plot). As the inspiral progresses the islands shift to lower $\rho$ coming closer to the central object located at $\rho=0$ and at the same time the eccentricity of the trajectory lowers. The depiction of the trapping is easier to see, when a \ix{\textit{stroboscopic} depiction} is employed on the section (right plot of Fig.~\ref{fig:res67}), i.e. from the time series of the left plot only every seventh consequent point is kept. The inset in the right panel of Fig.~\ref{fig:res67} shows the whole stroboscopic evolution of the inspiral on the section, while the main right panel focuses on the trapping in the islands of stability leaving out most of the "KAM phase" of the evolution. 

Let us take another look at the trajectories ``trapped'' in resonance. The initial condition giving the behavior seen in Fig.~\ref{fig:res67} is not unique, there is a range of them. For all of them the trapping lasts at least $10^4$ sections, which is the number of sections we allowed the inspiral to evolve to. Recall that this is the number of times the inspiral crosses the equatorial plane in a certain polar direction, so we can infer that $10^4$ is also roughly the total number of orbital cycles made by the particle. Since and entire EMRI takes $\sim 1/q$ orbital cycles, we then see that the particle has been trapped in the resonance for a time comparable with the entire inspiral time!

The quadrupole deviation of the spacetime, on which we have evolved the inspiral, is  $\Delta \mathcal{Q} \approx 23 M^3$. This number is extremely large, we expect that if there are any quadrupole deviations from Kerr, realistically they would be estimated to be of the order of $\mathcal{O}(10^{-4} M^3)$ \cite{Barausse20}. Even if the initial conditions giving such a prolonged resonance are several, they appear to be of zero measure in comparison with the initial conditions leading to the usual non-trapping resonances. There might be an issue also with the approximation we have employed to obtain this result. Namely, the linear approximation in energy and angular momentum make sense as long as the trajectory of the inspiral does not get too far from the initial geodesic, on which we have calculated the fluxes. This is not the case in our example. We have tried to address this issue by giving by hand fluxes deviating from the calculated ones up to at least the second significant digit in order to see if the trapping was a flux fine tuning effect. It turns out that this trapping of the inspiral does not appear to depend heavily on the specific values of the dissipative fluxes and the ratio $\dot{E}/\dot{L}_z$. However, it may be that the trapping occurs due to the fact that we are applying dissipation only to $E,L_Z$ and not to other components of angular momentum. Either way, this example would correspond to a \ix{\textit{sustained resonance}} as discussed in Section \ref{sec:resopass}.

\begin{figure}[ht] 
    \centering
    \includegraphics[width=0.49\textwidth]{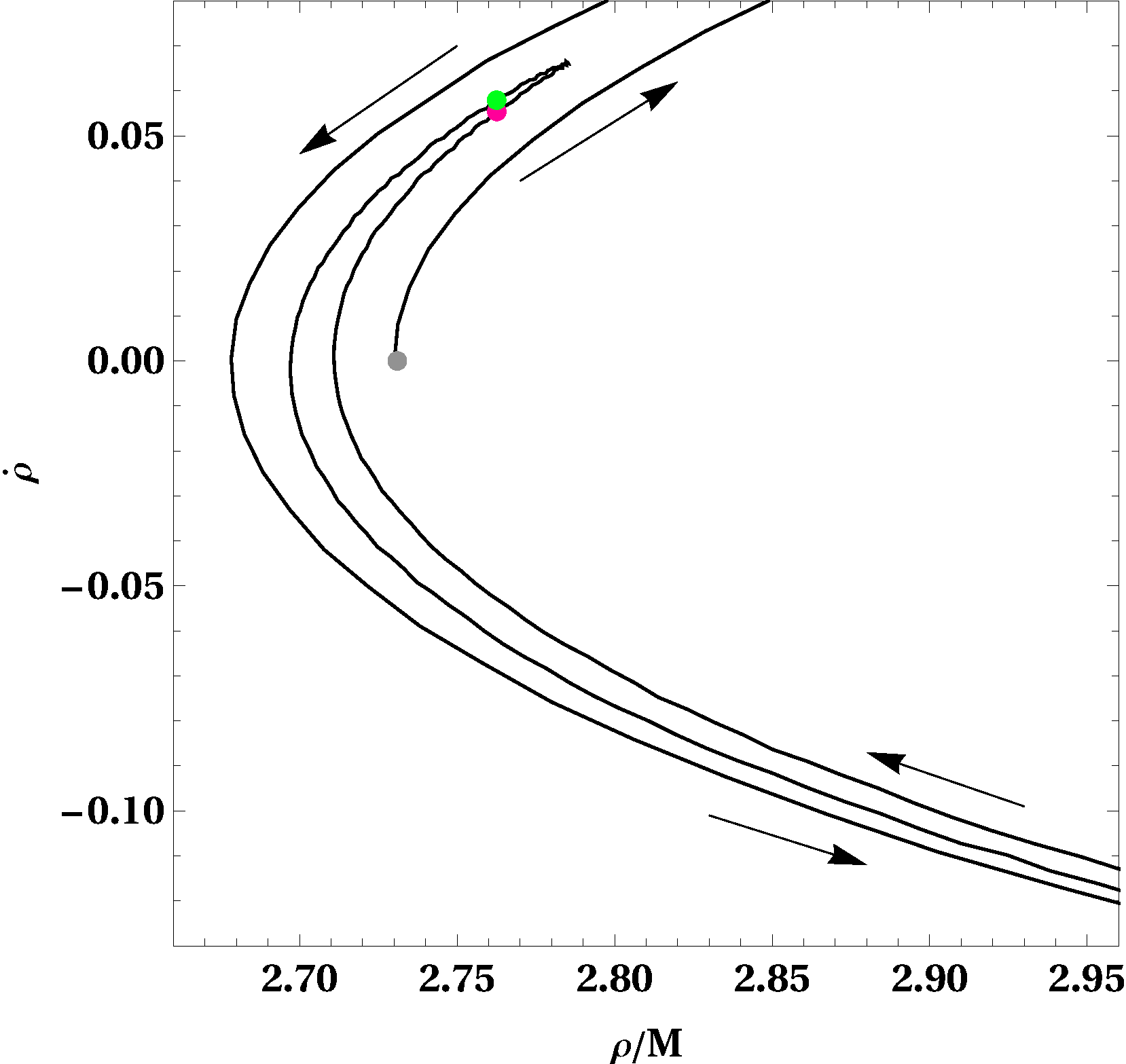}  
    \includegraphics[width=0.49\textwidth]{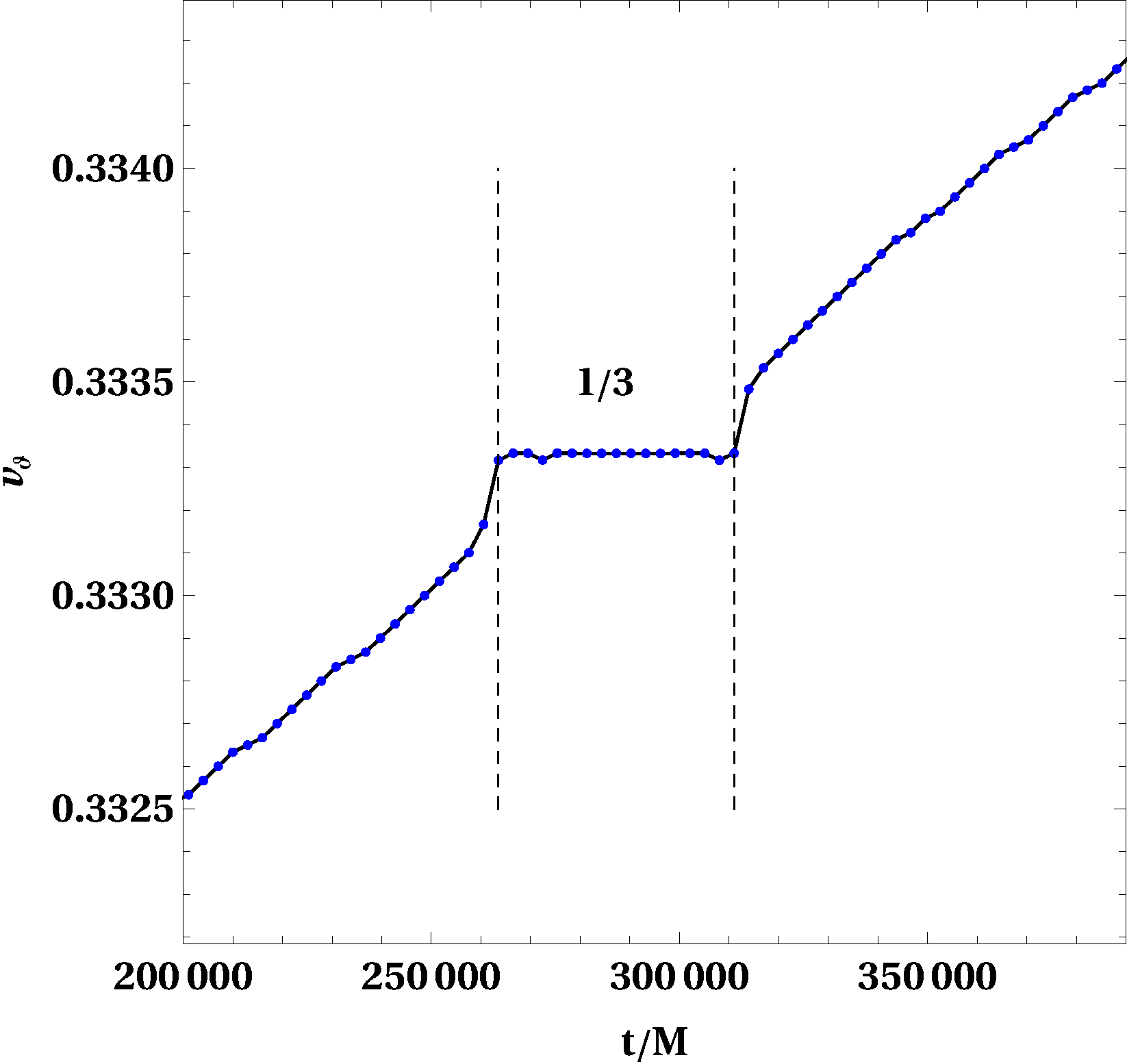}    
    \caption{An inspiral with mass ratio $q=10^{-5}$ crossing a $1/3$ resonance on a MSM spacetime with $a=0.999~M,~b=-4~10^{-4}~M$. Left panel: A stroboscopic depiction of the inspiral on a section lying on the equatorial plane with energy $E|_g=0.97 $ and angular momentun $L_z|_g=2 M$. The inspiral starts from the gray point $\rho=2.73~ M,~\dot{\rho}=0$, enters to the resonance at a position indicated by a red point and exits from the resonance at a point indicated by the green point. The arrows indicate the direction that the inspiral evolves. Right panel: The rotation curve corresponding to the trajectory shown in the left panel as a function of the coordinate time. The dashed line indicate the borders of the $1/3$ resonance. }
    \label{fig:res13}
\end{figure}

A more standard crossing corresponding to the transient-resonance scenario presented in \ref{sec:resopass} can be seen in our $1/3$ resonance crossing example. This has been calculated on a background with $\Delta \mathcal{Q}\approx 2\,10^{-4} M^3$ and with mass ratio set to $q=10^{-5}$. For presenting this example, we use only the stroboscopic depiction of a section (left panel of Fig.~\ref{fig:res13}) showing how the trajectory enters (red point) and leaves the resonance (green point). In this stroboscopic depiction the points belonging to the trajectory follow a clockwise direction until they reach the resonance; they follow the resonance as long as they are trapped in it, which is in this case is just a small part of an island of stability as the most cases during a crossing appear to be; once they leave the resonance, they follow a counterclockwise rotation. The crossing of the resonance lasted approximately for $650$ sections, which as explained previously corresponds to the number of cycles around the primary. When we compare it to the number of cycles expected from a transient resonance $\sqrt{\Delta \mathcal{Q}}/q\approx 1400$, we see that there are of the same order of magnitude. The number of cycles spent in this resonant represent a $\sim 1\%$ of the total expected cycles of a EMRI with $q=10^{-5}$. 

The actual information about how long did the inspiral stay in the resonance can be obtained by a \ix{rotation curve}. The right panel of Fig.~\ref{fig:res13} shows the rotation curve corresponding to the left panel, in which the horizontal axis is the coordinate time instead of initial condition shown in Fig.~\ref{fig:sec_rot}. There are two ways of producing this curve. The first is to apply Fourier analysis on the trajectory's time series to find the frequencies as done and described in detail in \cite{LGAC10}. However, this approach is cumbersome and the produced rotation curve at the resonance might give oscillations around the plateau as in \cite{LGAC10} instead of a clear plateau. An easier way, providing also a clearer plateau (right panel of Fig.~\ref{fig:res13}), is to take the points producing the curve shown in the left panel as initial conditions and evolve those points as geodesics for enough sections in order to produce the respective rotation curve (right panel of Fig.~\ref{fig:res13}). Note that by imposing the dissipation in our scheme has a result the non-conservation of the contraction of the  test body's  four-velocity. The initial value $(v_\mu v^\mu=-1)$ increases very slowly tending to zero, since the change is extremely slow the four-velocity remains time-like throughout the calculation. However, the actual values of the contraction have to be taken into account when reproducing the geodesics for the rotation curve.

\section{Discussion}

One of the other phenomena associated with resonances is the emergence of chaos. However, at the moment the impact of chaos on EMRIs and the corresponding gravitational waves is largely unexplored. Even if we do not expect chaos to have a significant impact on such systems, since the chaotic layers in phase space are very small under realistic perturbations, there might still be surprises around the corner. Some work has been done on the prolonged resonances \citep[e.g.][]{brink2015orbital}, but carrying out realistic computations of crossing such resonances was impossible until recently. The issue was that the self-force, as well as many idealized perturbations, only create axisymmetric perturbations to the motion, so the $\varphi$ coordinate stays redundant. As a result, resonances only arise in the $r, \vartheta$ sector, that is, for generically inclined and eccentric motion. However, both the dissipative fluxes and the self-force itself have been computed for the generic orbital case only quite recently \citep{van2018gravitational}. It is thus the next order of business to use this realistic self-force to study resonances in inspirals beyond the level of toy models such as the one presented in the past sections.

There has not yet been a consistent calculation or a systematic and exhaustive investigation of the crossings of resonances, even though these crossings, as shown in our examples, can last for a non-negligible number of cycles. The crossings will have a definite imprint on the emitted gravitational waves and the detection using matched filters, since the ratios of the frequencies of these waves will be rational for a larger number of cycles, thus modifying the frequency domain shape of the waveform template. In fact, since the impact of the resonance on the phase is $1/\sqrt{q}$, resonant effects have a \textit{higher} priority for inclusion than any other post-adiabatic effect. On the other hand, as discussed in Section \ref{sec:resopass}, we need the resonant phase $\Gamma$ at the beginning of the resonant crossing with sufficient accuracy to estimate the passage well, so the passage through non-resonant parts of the phase space still need to be computed at accuracy beyond adiabatic order. Alternatively, the resonant phase $\Gamma$ with which the inspiral enters the resonance can be understood as a free parameter of the waveform, even though that would reduce the predictivity and usefulness of the model.   

One of the strongest causes for studying resonances is the non-integrability the gravitational self-force itself can introduce into an inspiral. However, the prolonged resonances induced by the spin of the secondary \cite{Zelenka20} may provide another good case to study resonant effects. The numerical tools needed to calculate the fluxes of gravitational waves from the spinning particle such as numerical Teukolsky equations solvers will be available in the near future, if they are not already available. Taking into account only the first order dissipative part of the self force might not provide the whole picture of a prolonged-resonance crossing, but the main challenge at this stage is just to understand the phase-space features of a resonance leading to the different behaviors of the inspiral during the crossing. Ultimately, the adiabatic approximation of the inspiral will have to be replaced by a full self-force and \textit{self-torque} computation for an accurate picture.      

The analytical perturbative treatment sketched in Sec.~\ref{sec:resopass} should provide a more robust systematic framework to tackle the issue of passing through a resonance in semi-analytical inspiral models. However, this formalism currently only represents an order of magnitude estimate of the potential efficiency of a certain computation scheme. Nevertheless, it is yet to be shown whether and how such a scheme can work in practice and, in particular, what are the ``factors of order one'' in front of the leading order terms in the estimates in the $\epsilon \to 0$ limit. We leave this question for future work.

\section*{Acknowledgments}
GL-G has been supported by the fellowship Lumina Quaeruntur No. LQ100032102 of the Czech Academy of Sciences.  VW was supported by European Union’s Horizon 2020 research and innovation programme under grant agreement No 894881. The authors would like to thank Luk\'{a}\v{s} Polcar for allowing them to use the didactic example in section~\ref{sec:ext_matt}.

\bibliographystyle{apa_shortauthor.bst}
\bibliography{refs}

\begin{thebibliography}{}

\bibitem[\protect\astroncite{{Amaro-Seoane} et~al.}{2017}]{LISA}
{Amaro-Seoane}, P., {Audley}, H., {Babak}, S., et~al. (2017).
\newblock {Laser Interferometer Space Antenna}.
\newblock {\em arXiv:1702.00786}.

\bibitem[\protect\astroncite{Arnold et~al.}{2006}]{Arnold06}
Arnold, V., Kozlov, V., and Neishtadt, A. (2006).
\newblock {\em Mathematical Aspects of Classical and Celestial Mechanics}.
\newblock Springer International Publishing, 3rd edition.

\bibitem[\protect\astroncite{{Arnold}}{1963}]{Arnold63}
{Arnold}, V.~I. (1963).
\newblock {Proof of a Theorem of A. N. Kolmogorov on the Invariance of
  Quasi-Periodic Motions Under Small Perturbations of the Hamiltonian}.
\newblock {\em Russian Mathematical Surveys}, 18(5):9--36.

\bibitem[\protect\astroncite{{Bambi}}{2011}]{Bambi11}
{Bambi}, C. (2011).
\newblock {Testing the {K}err Black Hole Hypothesis}.
\newblock {\em Modern Physics Letters A}, 26(33):2453--2468.

\bibitem[\protect\astroncite{Banks et~al.}{1992}]{Banks92}
Banks, J., Brooks, J., Cairns, G., et~al. (1992).
\newblock On {D}evaney's definition of chaos.
\newblock {\em The American Mathematical Monthly}, 99(4):332--334.

\bibitem[\protect\astroncite{{Barack} and {Pound}}{2019}]{Barack19}
{Barack}, L. and {Pound}, A. (2019).
\newblock {Self-force and radiation reaction in general relativity}.
\newblock {\em Reports on Progress in Physics}, 82(1):016904.

\bibitem[\protect\astroncite{{Barausse} et~al.}{2020}]{Barausse20}
{Barausse}, E., {Berti}, E., {Hertog}, T., et~al. (2020).
\newblock {Prospects for fundamental physics with LISA}.
\newblock {\em General Relativity and Gravitation}, 52(8):81.

\bibitem[\protect\astroncite{Basovn{\'\i}k and
  Semer{\'a}k}{2016}]{basovnik2016geometry}
Basovn{\'\i}k, M. and Semer{\'a}k, O. (2016).
\newblock Geometry of deformed black holes. {II}. {S}chwarzschild hole
  surrounded by a {B}ach-{W}eyl ring.
\newblock {\em Physical Review D}, 94(4):044007.

\bibitem[\protect\astroncite{{Birkhoff}}{1913}]{Birkhoff13}
{Birkhoff}, G.~D. (1913).
\newblock {Proof of Poincar\'{e}'s geometric theorem}.
\newblock {\em Transactions of the American Mathematical Society},
  14(1):14--22.

\bibitem[\protect\astroncite{{Bičák} and {Ledvinka}}{1993}]{Bicak93}
{Bičák}, J. and {Ledvinka}, T. (1993).
\newblock {Relativistic Disks as sources of the {K}err metric}.
\newblock {\em Physical Review Letters}, 71(11):1669--1672.

\bibitem[\protect\astroncite{Brink et~al.}{2015}]{brink2015orbital}
Brink, J., Geyer, M., and Hinderer, T. (2015).
\newblock Orbital resonances around black holes.
\newblock {\em Physical review letters}, 114(8):081102.

\bibitem[\protect\astroncite{{Carter}}{1968a}]{Carter68}
{Carter}, B. (1968a).
\newblock {Global Structure of the {K}err Family of Gravitational Fields}.
\newblock {\em Physical Review}, 174(5):1559--1571.

\bibitem[\protect\astroncite{{Carter}}{1968b}]{Carter68b}
{Carter}, B. (1968b).
\newblock {{H}amilton-{J}acobi and {S}chrödinger Separable Solutions of
  {E}instein's {E}quations}.
\newblock {\em Communications in Mathematical Physics}, 10(4):280--310.

\bibitem[\protect\astroncite{Contopoulos}{2004}]{contopoulos2004}
Contopoulos, G. (2004).
\newblock {\em Order and chaos in dynamical astronomy}.
\newblock Springer Science \& Business Media.

\bibitem[\protect\astroncite{{Dixon}}{1974}]{Dixon74}
{Dixon}, W.~G. (1974).
\newblock {Dynamics of Extended Bodies in {G}eneral {R}elativity. {III.}
  {E}quations of Motion}.
\newblock {\em Philosophical Transactions of the Royal Society of London Series
  A}, 277(1264):59--119.

\bibitem[\protect\astroncite{Doroshkevich et~al.}{1966}]{doroshkevich1966}
Doroshkevich, A., Zel’Dovich, Y.~B., and Novikov, I. (1966).
\newblock Gravitational collapse of nonsymmetric and rotating masses.
\newblock {\em Sov. Phys.—JETP}, 22:122--30.

\bibitem[\protect\astroncite{{Efthymiopoulos} et~al.}{1997}]{Efthymiopoulos97}
{Efthymiopoulos}, C., {Contopoulos}, G., {Voglis}, N., and {Dvorak}, R. (1997).
\newblock {Stickiness and cantori}.
\newblock {\em Journal of Physics A Mathematical General}, 30(23):8167--8186.

\bibitem[\protect\astroncite{{Eleni} and {Apostolatos}}{2020}]{Eleni20}
{Eleni}, A. and {Apostolatos}, T.~A. (2020).
\newblock {Newtonian analogue of a {K}err black hole}.
\newblock {\em Physical Review D}, 101(4):044056.

\bibitem[\protect\astroncite{{Flanagan} and {Hinderer}}{2012}]{Flanagan12}
{Flanagan}, {\'E}.~{\'E}. and {Hinderer}, T. (2012).
\newblock {Transient Resonances in the Inspirals of Point Particles into Black
  Holes}.
\newblock {\em Physical Review Letters}, 109(7):071102.

\bibitem[\protect\astroncite{{Frolov} et~al.}{2017}]{Frolov17}
{Frolov}, V.~P., {Krtou{\v{s}}}, P., and {Kubiz{\v{n}}{\'a}k}, D. (2017).
\newblock {Black holes, hidden symmetries, and complete integrability}.
\newblock {\em Living Reviews in Relativity}, 20(1):6.

\bibitem[\protect\astroncite{{Grobman}}{1959}]{Grobman59}
{Grobman}, D.~M. (1959).
\newblock {Homeomorphisms of systems of differential equations}.
\newblock {\em Doklady Akademii Nauk SSSR}, 128:880--881.

\bibitem[\protect\astroncite{Hansen}{1974}]{hansen1974}
Hansen, R.~O. (1974).
\newblock Multipole moments of stationary space-times.
\newblock {\em Journal of Mathematical Physics}, 15(1):46--52.

\bibitem[\protect\astroncite{{Hartman}}{1960}]{Hartman60}
{Hartman}, P. (1960).
\newblock {A lemma in the theory of structural stability of differential
  equations}.
\newblock {\em Proceedings of the American Mathematical Society},
  11(4):610--620.

\bibitem[\protect\astroncite{Isoyama et~al.}{2013}]{isoyama2013}
Isoyama, S., Fujita, R., Nakano, H., et~al. (2013).
\newblock Evolution of the carter constant for resonant inspirals into a {K}err
  black hole: {I.}the scalar case.
\newblock {\em Progress of Theoretical and Experimental Physics},
  2013(6):063E01.

\bibitem[\protect\astroncite{Isoyama et~al.}{2019}]{isoyama2019}
Isoyama, S., Fujita, R., Nakano, H., et~al. (2019).
\newblock “flux-balance formulae” for extreme mass-ratio inspirals.
\newblock {\em Progress of Theoretical and Experimental Physics},
  2019(1):013E01.

\bibitem[\protect\astroncite{{Johannsen}}{2013}]{Johannsen13}
{Johannsen}, T. (2013).
\newblock {Regular black hole metric with three constants of motion}.
\newblock {\em Physical Review D}, 88(4):044002.

\bibitem[\protect\astroncite{{Kerr}}{1963}]{Kerr63}
{Kerr}, R.~P. (1963).
\newblock {Gravitational Field of a Spinning Mass as an Example of
  Algebraically Special Metrics}.
\newblock {\em Physical Review Letters}, 11(5):237--238.

\bibitem[\protect\astroncite{Kevorkian and Cole}{2012}]{Kevorkian12}
Kevorkian, J.~K. and Cole, J.~D. (2012).
\newblock {\em Multiple scale and singular perturbation methods}, volume 114.
\newblock Springer Science \& Business Media.

\bibitem[\protect\astroncite{{Kolmogorov}}{1954}]{Kolmogorov54}
{Kolmogorov}, A.~N. (1954).
\newblock {On the Conservation of Conditionally Periodic Motions under Small
  Perturbation of the Hamiltonian}.
\newblock {\em Doklady Akademii Nauk SSSR}, 98:527--530.

\bibitem[\protect\astroncite{{Lemos} and {Letelier}}{1994}]{Lemos94}
{Lemos}, J. P.~S. and {Letelier}, P.~S. (1994).
\newblock {Exact general relativistic thin disks around black holes}.
\newblock {\em Physical Review D}, 49(10):5135--5143.

\bibitem[\protect\astroncite{Lukes-Gerakopoulos et~al.}{2010}]{LGAC10}
Lukes-Gerakopoulos, G., Apostolatos, T.~A., and Contopoulos, G. (2010).
\newblock {Observable signature of a background deviating from the {K}err
  metric}.
\newblock {\em Phys. Rev. D}, 81:124005.

\bibitem[\protect\astroncite{{Lukes-Gerakopoulos} et~al.}{2016}]{LG16}
{Lukes-Gerakopoulos}, G., {Katsanikas}, M., {Patsis}, P.~A., and {Seyrich}, J.
  (2016).
\newblock {Dynamics of a spinning particle in a linear in spin Hamiltonian
  approximation}.
\newblock {\em Physical Review D}, 94(2):024024.

\bibitem[\protect\astroncite{Lynden-Bell}{2000}]{lynden2000carter}
Lynden-Bell, D. (2000).
\newblock Carter separable electromagnetic fields.
\newblock {\em Monthly Notices of the Royal Astronomical Society},
  312(2):301--315.

\bibitem[\protect\astroncite{{Manko} and {Novikov}}{1992}]{Manko92}
{Manko}, V.~S. and {Novikov}, I.~D. (1992).
\newblock {Generalizations of the {K}err and {K}err-{N}ewman metrics possessing
  an arbitrary set of mass-multipole moments}.
\newblock {\em Classical and Quantum Gravity}, 9(11):2477--2487.

\bibitem[\protect\astroncite{{Manko} et~al.}{2000}]{MSM}
{Manko}, V.~S., {Sanabria-G{\'o}mez}, J.~D., and {Manko}, O.~V. (2000).
\newblock {Nine-parameter electrovac metric involving rational functions}.
\newblock {\em Physical Review D}, 62(4):044048.

\bibitem[\protect\astroncite{{Markakis}}{2014}]{Markakis14}
{Markakis}, C. (2014).
\newblock {Constants of motion in stationary axisymmetric gravitational
  fields}.
\newblock {\em Monthly Notices of the Royal Astronomical Society},
  441(4):2974--2985.

\bibitem[\protect\astroncite{Mathisson}{1937}]{Mathisson37}
Mathisson, M. (1937).
\newblock {Neue mechanik materieller systemes}.
\newblock {\em Acta Phys. Polon.}, 6:163--2900.

\bibitem[\protect\astroncite{Miller and Pound}{2020}]{Miller20}
Miller, J. and Pound, A. (2020).
\newblock Two-timescale evolution of extreme-mass-ratio inspirals: waveform
  generation scheme for quasicircular orbits in {S}chwarzschild spacetime.
\newblock {\em arXiv preprint arXiv:2006.11263}.

\bibitem[\protect\astroncite{{Mino}}{2003}]{Mino03}
{Mino}, Y. (2003).
\newblock {Perturbative approach to an orbital evolution around a supermassive
  black hole}.
\newblock {\em Physical Review D}, 67(8):084027.

\bibitem[\protect\astroncite{Morbidelli}{2002}]{Morbidelli02}
Morbidelli, A. (2002).
\newblock {\em Modern celestial mechanics: aspects of solar system dynamics}.
\newblock CRC Press, 1st edition.

\bibitem[\protect\astroncite{{Moser}}{1962}]{Moser62}
{Moser}, J. (1962).
\newblock {On invariant curves of area-preserving mappings of an annulus}.
\newblock {\em Nachrichten der Akademie der Wissenschaften in G\"ottingen.
  {II}. Mathematisch-Physikalische Klasse}, pages 1--20.

\bibitem[\protect\astroncite{{Neugebauer} and {Meinel}}{1993}]{Neugebauer93}
{Neugebauer}, G. and {Meinel}, R. (1993).
\newblock {The {E}insteinian Gravitational Field of the Rigidly Rotating Disk
  of Dust}.
\newblock {\em Astrophysical Journal Letters}, 414:L97.

\bibitem[\protect\astroncite{Papapetrou}{1951}]{Papapetrou51}
Papapetrou, A. (1951).
\newblock {Spinning test particles in general relativity. 1.}
\newblock {\em Proc. Roy. Soc. Lond.}, A209:248--258.

\bibitem[\protect\astroncite{{Pesin}}{1977}]{Pesin77}
{Pesin}, Y.~B. (1977).
\newblock {Characteristic Lyapunov Exponents and Smooth Ergodic Theory}.
\newblock {\em Russian Mathematical Surveys}, 32(4):55--114.

\bibitem[\protect\astroncite{{Poincar\'{e}}}{1912}]{Poincare12}
{Poincar\'{e}}, H. (1912).
\newblock {Sur un th\'{e}or\`{e}me de g\'{e}om\'{e}trie}.
\newblock {\em {Rendiconti del Circolo Matematico di Palermo}}, 33:375--407.

\bibitem[\protect\astroncite{{Poincar\'{e}}}{1993}]{Poincare93}
{Poincar\'{e}}, H. (1993).
\newblock {\em {New methods of celestial mechanics}}.
\newblock American Institute of Physics, Woodbury, NY, New York City.

\bibitem[\protect\astroncite{Poisson et~al.}{2011}]{poisson2011}
Poisson, E., Pound, A., and Vega, I. (2011).
\newblock The motion of point particles in curved spacetime.
\newblock {\em Living Reviews in Relativity}, 14(1):7.

\bibitem[\protect\astroncite{Polcar and Semer{\'a}k}{2019}]{polcar2019free}
Polcar, L. and Semer{\'a}k, O. (2019).
\newblock Free motion around black holes with discs or rings: Between
  integrability and chaos. {VI.} the melnikov method.
\newblock {\em Physical Review D}, 100(10):103013.

\bibitem[\protect\astroncite{R{\"u}diger}{1981}]{ruediger1}
R{\"u}diger, R. (1981).
\newblock Conserved quantities of spinning test particles in general
  relativity. i.
\newblock {\em Proc. Royal Soc. Lond. A}, 375(1761):185--193.

\bibitem[\protect\astroncite{R{\"u}diger}{1983}]{ruediger2}
R{\"u}diger, R. (1983).
\newblock Conserved quantities of spinning test particles in general
  relativity. {II}.
\newblock {\em Proc. Royal Soc. Lond. A}, 385(1788):229--239.

\bibitem[\protect\astroncite{Sano and Tagoshi}{2014}]{sano2014gravitational}
Sano, Y. and Tagoshi, H. (2014).
\newblock Gravitational perturbation induced by a rotating ring around a {K}err
  black hole.
\newblock {\em arXiv preprint arXiv:1412.8607}.

\bibitem[\protect\astroncite{{Semer{\'a}k}}{2003}]{Semerak03}
{Semer{\'a}k}, O. (2003).
\newblock {Gravitating discs around a {S}chwarzschild black hole: {III}}.
\newblock {\em Classical and Quantum Gravity}, 20(9):1613--1634.

\bibitem[\protect\astroncite{Semer{\'a}k and Sukov{\'a}}{2010}]{semerak2010}
Semer{\'a}k, O. and Sukov{\'a}, P. (2010).
\newblock Free motion around black holes with discs or rings: between
  integrability and chaos--i.
\newblock {\em Monthly Notices of the Royal Astronomical Society},
  404(2):545--574.

\bibitem[\protect\astroncite{{Semer{\'a}k} and
  {{\v{C}}{\'\i}{\v{z}}ek}}{2020}]{Semerak20}
{Semer{\'a}k}, O. and {{\v{C}}{\'\i}{\v{z}}ek}, P. (2020).
\newblock {Rotating Disc around a {S}chwarzschild Black Hole}.
\newblock {\em Universe}, 6(2):27.

\bibitem[\protect\astroncite{Semerák and Suková}{2015}]{Semerak15}
Semerák, O. and Suková, P. (2015).
\newblock {On Geodesic Dynamics in Deformed Black-Hole Fields}.
\newblock {\em Fund. Theor. Phys.}, 179:561--586.

\bibitem[\protect\astroncite{Silverman}{1992}]{silverman1992maps}
Silverman, S. (1992).
\newblock On maps with dense orbits and the definition of chaos.
\newblock {\em The Rocky Mountain Journal of Mathematics}, 22(1):353--375.

\bibitem[\protect\astroncite{Smale}{1965}]{smale1965}
Smale, S. (1965).
\newblock Diffeomorphisms with many periodic points.
\newblock In Cairns, S.~S., editor, {\em Differential and combinatorial
  topology: a symposium in honor of Marston Morse}. Princeton University Press.

\bibitem[\protect\astroncite{{Suzuki} and {Maeda}}{1997}]{Suzuki97}
{Suzuki}, S. and {Maeda}, K.-I. (1997).
\newblock {Chaos in {S}chwarzschild spacetime: The motion of a spinning
  particle}.
\newblock {\em Physical Review D}, 55(8):4848--4859.

\bibitem[\protect\astroncite{{van de Meent}}{2014}]{vandeMeent14}
{van de Meent}, M. (2014).
\newblock {Conditions for sustained orbital resonances in extreme mass ratio
  inspirals}.
\newblock {\em Physical Review D}, 89(8):084033.

\bibitem[\protect\astroncite{van~de Meent}{2014}]{maarten2014}
van~de Meent, M. (2014).
\newblock Resonantly enhanced kicks from equatorial small mass-ratio inspirals.
\newblock {\em Physical Review D}, 90(4):044027.

\bibitem[\protect\astroncite{Van De~Meent}{2018}]{van2018gravitational}
Van De~Meent, M. (2018).
\newblock Gravitational self-force on generic bound geodesics in {K}err
  spacetime.
\newblock {\em Physical Review D}, 97(10):104033.

\bibitem[\protect\astroncite{{Vigeland} et~al.}{2011}]{Vigeland11}
{Vigeland}, S., {Yunes}, N., and {Stein}, L.~C. (2011).
\newblock {Bumpy black holes in alternative theories of gravity}.
\newblock {\em Physical Review D}, 83(10):104027.

\bibitem[\protect\astroncite{{Witzany}}{2019}]{Witzany19b}
{Witzany}, V. (2019).
\newblock {Hamilton-Jacobi equation for spinning particles near black holes}.
\newblock {\em Physical Review D}, 100(10):104030.

\bibitem[\protect\astroncite{Witzany et~al.}{2015}]{witzany2015}
Witzany, V., Semer{\'a}k, O., and Sukov{\'a}, P. (2015).
\newblock Free motion around black holes with discs or rings: between
  integrability and chaos--iv.
\newblock {\em Monthly Notices of the Royal Astronomical Society},
  451(2):1770--1794.

\bibitem[\protect\astroncite{{Witzany} et~al.}{2019}]{Witzany19}
{Witzany}, V., {Steinhoff}, J., and {Lukes-Gerakopoulos}, G. (2019).
\newblock {Hamiltonians and canonical coordinates for spinning particles in
  curved space-time}.
\newblock {\em Classical and Quantum Gravity}, 36(7):075003.

\bibitem[\protect\astroncite{{Zelenka} et~al.}{2020}]{Zelenka20}
{Zelenka}, O., {Lukes-Gerakopoulos}, G., {Witzany}, V., and
  {Kop{\'a}{\v{c}}ek}, O. (2020).
\newblock {Growth of resonances and chaos for a spinning test particle in the
  {S}chwarzschild background}.
\newblock {\em Physical Review D}, 101(2):024037.

\end{thebibliography}

\end{document}